\documentclass[aps,twocolumn,superscriptaddress,longbibliography,nofootinbib]{revtex4-2}

\usepackage{amsmath,amssymb}
\usepackage{graphicx}
\usepackage{booktabs}
\usepackage[colorlinks=true,linkcolor=blue,citecolor=blue,urlcolor=blue]{hyperref}
\usepackage{xcolor}
\makeatletter
\def\paragraph{%
  \@startsection{paragraph}{4}{\parindent}%
    {1.2ex plus .4ex minus .2ex}{-1em}{\normalfont\normalsize\itshape}}
\makeatother

\newcommand{\ket}[1]{\lvert #1 \rangle}
\newcommand{\Mhh}{M_{HH}}
\newcommand{\Mhv}{M_{HV}}
\newcommand{\Mvh}{M_{VH}}
\newcommand{\Mvv}{M_{VV}}
\newcommand{\purity}{\mathcal{C}}   %
\newcommand{\Phz}{\Phi_0}
\newcommand{\dT}{\Delta_\tau}
\newcommand{\dephw}{(M^\dagger M)_{HV}}

\begin{document}

\title{Single-shot coherent process tomography and mid-infrared polarimetry\\
with undetected photons}

\author{Marthe Zeja}
\email{marthe.zeja@physik.hu-berlin.de} %
\affiliation{Institut f\"ur Physik, Humboldt-Universit\"at zu Berlin, Newtonstra{\ss}e 15, 12489 Berlin, Germany}

\author{Wen-Zhe Yan}
\affiliation{CAS Key Laboratory of Quantum Information, University of Science and Technology of China, Hefei 230026, China}
\affiliation{CAS Center For Excellence in Quantum Information and Quantum Physics, University of Science and Technology of China, Hefei 230026, China}

\author{Zhibo Hou}
\affiliation{CAS Key Laboratory of Quantum Information, University of Science and Technology of China, Hefei 230026, China}
\affiliation{CAS Center For Excellence in Quantum Information and Quantum Physics, University of Science and Technology of China, Hefei 230026, China}
\affiliation{Hefei National Laboratory, University of Science and Technology of China, Hefei 230088, China}

\author{Sven Ramelow}
\affiliation{Institut f\"ur Physik, Humboldt-Universit\"at zu Berlin, Newtonstra{\ss}e 15, 12489 Berlin, Germany}
\affiliation{Ferdinand-Braun-Institut, Leibniz-Institut f\"ur H\"ochstfrequenztechnik, Gustav-Kirchhoff-Stra{\ss}e 4, 12489 Berlin, Germany}

\date{\today}

\begin{abstract}
Measurements with undetected photons infer properties of a probe beam that is
never detected, transferring mid-infrared information onto silicon-friendly
wavelengths. Tomography in this paradigm has so far relied on switched settings or scanned
phases, referenced across a drift-sensitive campaign; no protocol reads all
Jones parameters from one frame. Here we show that a folded nonlinear interferometer with a
diagonally pumped crossed-crystal source performs \emph{coherent
process tomography of the undetected beam in a single spectrometer frame}: it
reconstructs all seven parameters of the beam's round-trip Jones matrix, the
coherent part of the channel.
Birefringent group-delay walk-off, normally a calibration nuisance,
acts as a frequency multiplexer: each Jones-matrix element is assigned its own
carrier, a spectral fringe period paired with a detector port. Because the
multiplexing exploits the down-conversion bandwidth rather than merely
tolerating it, a single Fourier transform per detector trace returns all four
moduli and all three relative phases across that bandwidth, i.e., the full
round-trip Jones spectrum $M(\lambda)$ up to a global phase, with no scan or
setting change (after a one-time reference frame).
Simultaneity makes every relative phase immune to common-mode drift, and a
dark carrier, a self-interference term that unitarity forces to vanish,
provides a built-in null test. For realistic parameters of a
periodically poled KTP source the scheme yields wavelength-resolved
mid-infrared polarimetry across $3.5$--$4.2\,\mu$m (${\sim}80$ spectral
points per Jones element) on a silicon camera,
extending to depth-resolved Jones matrices, i.e.,\ polarization-sensitive
optical coherence tomography with undetected photons. A Fisher-information
analysis, the first we are aware of in undetected-photon tomography, sets the
precision budget and compares delay multiplexing to sequential protocols at
equal photon number.
\end{abstract}

\maketitle

\section{Introduction}
\label{sec:intro}
\begin{figure*}[t]
\includegraphics[width=\textwidth]{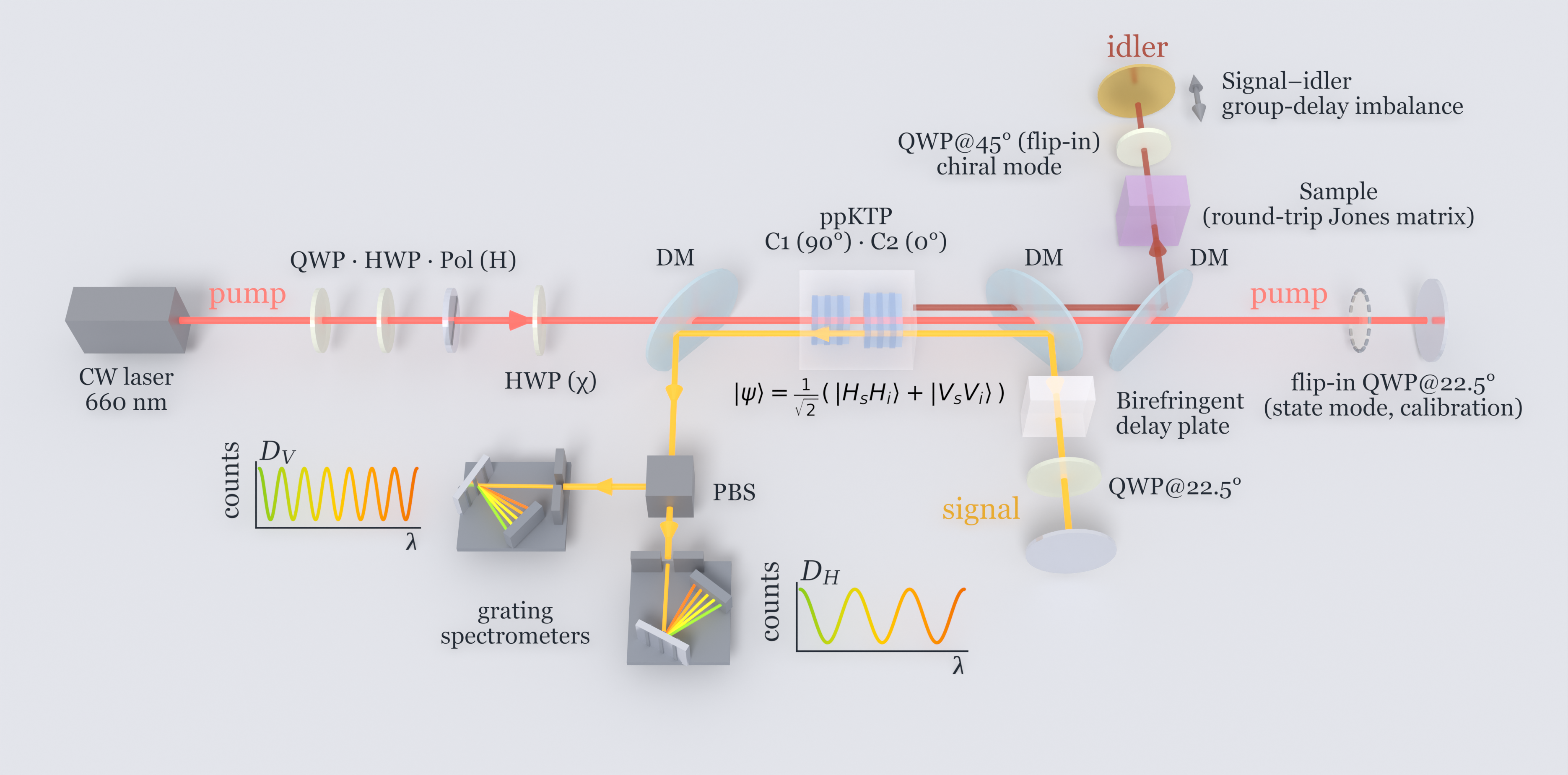}
\caption{\label{fig:setup}%
Folded nonlinear interferometer for single-shot coherent process tomography.
A continuous-wave laser at $660$\,nm is prepared in $H$ by a quarter-wave
plate, a half-wave plate and a polarizer; a further half-wave plate sets the
pump angle $\chi$, which equals $45^\circ$ at the balanced working point
(Sec.~\ref{sec:working}). The crossed type-0 crystals (periodically poled
KTiOPO$_4$, ppKTP) generate
signal--idler pairs by parametric down-conversion and are double-passed: the pump
meets C1 ($90^\circ$, generating the $V$ pair) first and C2 ($0^\circ$, generating
the $H$ pair) second; see App.~\ref{app:rates}. The state label is the first-pass
pair at the balanced point $\chi = 45^\circ$ [Eq.~(\ref{eq:pass1}), with $\kappa$
absorbed into the calibration constants of App.~\ref{app:rates}]. Dichroic mirrors
split pump, signal, and idler into three arms with end mirrors. The signal arm
carries the birefringent delay plate $\mathcal{P}(T)$, which sets the carrier splitting
$d_1$ of the family ladder (Sec.~\ref{sec:walkoff}), and a quarter-wave plate at
$22.5^\circ$ acting in double pass as the reflection $R(A)$ (Sec.~\ref{sec:rates})
with $A = 45^\circ$, twice the plate axis angle. The idler arm holds the unknown channel,
whose round-trip Jones matrix $M$ is to be determined; a quarter-wave plate at
$45^\circ$ is flipped in only for the chiral operating mode
(Sec.~\ref{sec:discussion}). The pump arm holds a bare end mirror and a flip-in
quarter-wave plate at $22.5^\circ$, used in the $\chi = 0$ state mode (it returns the
pump at the polarization angle $B = A$, Sec.~\ref{sec:diagonal}) and as a calibration switch
(App.~\ref{app:transits}). On the return pass
the pump generates pairs in both crystals a second time; the returning signal is split on a
polarizing beam splitter (PBS) into its $H$ and $V$ components, whose two ports
$D_H$ and $D_V$ feed two grating
spectrometers; the sketched traces are the superposed spectral fringes that
carry the Jones matrix. The double arrow marks $\dT$, the fixed signal--idler
group-delay imbalance that maps interferometer phase onto those fringes.
Abbreviations used in the figure: QWP/HWP, quarter-/half-wave plate; Pol,
polarizer; CW, continuous wave; DM, dichroic mirror; PBS, polarizing beam
splitter.}
\end{figure*}

Nonlinear interferometers with undetected photons
\cite{lemos2014,chekhova2016b,hochrainer2022,lemos2022} measure a sample with light that is never
detected. Rooted in the SU(1,1) interferometer of Yurke et al.\ \cite{yurke1986}
and the induced-coherence experiments of Zou, Wang, and Mandel
\cite{zou1991,wangzoumandel1991},
they exploit path identity to transfer amplitude
and phase information from an undetected idler, typically in the mid-infrared
(mid-IR), onto a detected signal at a convenient wavelength. The paradigm has produced imaging and microscopy
\cite{lemos2014,kviatkovsky2020}, hyperspectral imaging
\cite{paterova2020,placke2025}, spectroscopy
\cite{kalashnikov2016,lindner2019,mukai2021,kaufmann2022}, and optical
coherence tomography (OCT)
\cite{paterova2017,vanselow2020,machado2020,kim2023}, with dispersion analysis and
compensation \cite{zorin2026} and phase-quadrature and holographic single-frame
readout \cite{haase2022,pearce2024,leontorres2024,topfer2024} extending
the toolbox.
Nearly all of these experiments operate in the low-gain regime of spontaneous
pair generation, which needs only a continuous-wave pump and probes the sample
at the vanishing intensity of spontaneous emission; each detected event then
interferes single-pair amplitudes. We adopt this regime throughout.
High-parametric-gain operation, demonstrated for spectroscopy
\cite{hashimoto2023,hashimoto2024} and low-coherence interferometry
\cite{machado2020,zotti2025}, and analyzed in Ref.~\cite{houde2026b},
brightens the beams at the price of a different noise model and is deferred to
the outlook (Sec.~\ref{sec:discussion}).

Recently the paradigm has been pushed toward \emph{tomography}: polarization state tomography
of the undetected photon \cite{fuenzalida2022b}, polarimetry of
birefringent and diattenuating samples \cite{paterova2018b,oglialoro2025}, the full
Jones matrix from wave-plate settings in a two-source geometry
\cite{arya2025b}, and quantum process tomography, proposed for unitary operations
\cite{rajeev2026b} and, in a single-Kraus
form, for general linear ones acting on the undetected beam \cite{rajeev2026}. All of these protocols share one architecture: information is
extracted \emph{sequentially}, frame by frame. Most of them switch known wave-plate
or mode-unitary settings between interferograms
\cite{fuenzalida2022b,paterova2018b,rajeev2026b,rajeev2026}; Ref.~\cite{oglialoro2025}
instead ramps two phases at the same time and separates them in the Fourier domain
of the scan. Either way, fringe phases are referenced to a common phase standard
maintained across the campaign. This costs time linearly in the number of settings or scan steps and,
more fundamentally, exposes every phase difference to interferometer drift between
frames, a limitation that the proponents of these schemes state themselves:
Ref.~\cite{paterova2018b} requires the interferometer to be well stabilized in its
phase-shift method for the duration of the measurements, and Ref.~\cite{oglialoro2025} notes that the
uncertainties reported in that experiment were dominated by phase drifts. A second gap is conceptual: none of these tomography papers quantifies the
attainable precision, so there is no Fisher information, no Cram\'er--Rao bound
(CRB) benchmark, and no design theory.
The gap is not for lack of tools; rather, no field has combined them. Classical
polarimetry has the estimation theory: Stokes and Mueller polarimeters have long
been designed by their CRBs and estimator variances
\cite{sabatke2000,tyo2002,twietmeyer2008,roux2005,goudail2017,foreman2019}, but for
detected light and for Stokes, Mueller, or ellipsometric parameters, never for a full
complex Jones matrix.
Jones-matrix OCT has the object, an interferometrically measured Jones matrix, but
its precision analysis stops at bias-corrected and maximum-likelihood estimators
for the scalar retardance \cite{duan2011,kasaragod2014}, and, to our knowledge, CRB
analyses in OCT have addressed only scalar quantities, Doppler frequency and
attenuation \cite{yazdanfar2005,neubrand2022}. The
undetected-photon paradigm has neither: quantum estimation theory has so far
treated only one or two real parameters per sample: the transmission and phase
\cite{panda2026}, the absorption across source architectures \cite{houde2026},
and a joint absorption--dispersion estimate \cite{zhang2025h}.

Here we close both gaps at once: the sequential architecture and the missing
estimation theory. We fold the crossed-crystal source of
Ref.~\cite{fuenzalida2022b} into a Michelson-type nonlinear interferometer and pump it
\emph{diagonally}, so that the first-pass photon pair is polarization entangled and the
undetected idler probes the unknown channel $M$ (a general round-trip Jones matrix,
losses included) with both basis polarizations at once, as in
Refs.~\cite{oglialoro2025,fuenzalida2022b}. Two things are new. First, the reference
pairs of the return pass carry both polarizations as well, because both crystals emit
on both passes, so all four Jones elements interfere in the same frame. Second, the
four interference terms are separated not by settings but by delay: the
birefringent group-delay walk-off of the crossed crystals --- ordinarily a nuisance to
be compensated --- assigns each interfering amplitude pair its own carrier, a
spectral fringe period in a given detector port.
The two spectrometer traces therefore carry four bright
fringe \emph{families}, superposed in each trace but distinguished by their
fringe periods (two of them share a period and are told apart by the port
alone), plus a fifth, provably dark carrier. A single Fourier transform
per trace, that is, one spectrometer frame, separates them into distinct
delay-space peaks and yields amplitude and phase of every family
simultaneously: $4\times 2 - 1 = 7$ observables,
exactly the parameter count of a Jones matrix modulo an unobservable global phase.
Delay multiplexing of polarization channels has classical ancestors in Jones-matrix
OCT \cite{baumann2012,lim2012,ju2013,braaf2014}; a parallel-detection
alternative avoids the multiplexing entirely \cite{yamanari2019}. Transplanting it
into the undetected-photon paradigm is what turns it into single-shot
\emph{coherent} process tomography of a beam that is never detected. By coherent
process tomography we mean the reconstruction of the coherent, single-Kraus part of
the channel, its Jones matrix $M$, with all seven of its parameters; the
decoherent part of a general channel is not resolved within one frame by any low-gain
undetected-photon scheme (Sec.~\ref{sec:discussion}).

Four features distinguish the scheme from its sequential counterparts.
(i)~\emph{Single-shot, drift-immune readout}: all seven parameters are read from one
frame, relative phases are instantaneous, common-mode arm drift cancels, and a
dark carrier free of the global interferometer phase $\Phz$, forced to vanish
for every physical channel by
unitarity, gives a built-in null test (Sec.~\ref{sec:readout}).
(ii)~\emph{Broadband, spectrally resolved output}: the
scheme does not merely tolerate the down-conversion bandwidth; it runs on it --- the
delay carriers exist only across a broad spectrum, and demodulating each family returns
the full complex Jones \emph{spectrum} $M(\lambda)$ rather than one matrix per
campaign. That is about $80$ spectral points per element (Sec.~\ref{sec:budget}), whereas the
sequential protocols reconstruct a single parameter set per campaign. (iii)~\emph{Mid-IR reach}:
for realistic parameters of a periodically poled KTiOPO$_4$ (ppKTP) source,
this becomes wavelength-resolved mid-IR polarimetry and, for layered samples,
depth-resolved Jones-matrix polarimetry, i.e.,\ polarization-sensitive OCT (PS-OCT)
(Secs.~\ref{sec:feasibility} and \ref{sec:discussion}).
(iv)~\emph{Estimation theory}: we provide the first Fisher-information analysis, to
our knowledge, in the undetected-photon tomography literature, covering the
sequential-versus-multiplexed comparison at fixed photon budget and the working-point
design problem (Sec.~\ref{sec:fim}), a quantum-Fisher hierarchy that bounds the
price of not detecting the idler (Sec.~\ref{sec:qfi}), and a Monte Carlo
demonstration that the single-frame readout attains the classical bound
(Sec.~\ref{sec:noise}).
Table~\ref{tab:priorart} locates the scheme in the prior art.

\begin{table*}[t]
\caption{\label{tab:priorart}%
Prior work closest to the present scheme, feature by feature. ``Und.'': the sample
is probed by photons that are never detected. ``Single-shot'': all parameters are
read from one frame without switching settings; a sample-free reference measurement
and lateral scanning do not count as switching, tuning beyond the band of one frame
does. ``JM'': the full complex transfer matrix of the channel, including the axis
orientation in the polarization case ($2\times2$).
``$\lambda$'': the parameters are returned as a function of wavelength within the
same frame, at a resolution that may be coarser than the spectrometer's (about $80$
points here, Sec.~\ref{sec:budget}). ``Depth'':
depth-resolved in the sample. ``FIM'': Fisher-information, Cram\'er--Rao or
optimal-design analysis. (\checkmark) marks a partial entry. Under JM it means a
single retardance or a restricted model \cite{paterova2018b,oglialoro2025}, a Faraday
rotation covering $600$\,nm of idler wavelength by tuning the source one wavelength at
a time \cite{chakraborty2025}, one column of the matrix of a vortex wave plate from
two holograms \cite{vasikonis2026}, a general matrix proposed but not yet demonstrated
\cite{rajeev2026b,rajeev2026,arya2025b}, the Stokes
parameters of a beam \cite{oka1999} and Mueller elements
\cite{hagen2007}, demonstrated experimentally in
Refs.~\cite{dubreuil2007,hagen2021}, which fix the Jones matrix only for a
non-depolarizing sample. In the Jones-matrix OCT row the full matrix rests on
Refs.~\cite{lim2012,braaf2014}; Ref.~\cite{jones2026} returns retardance and axis
alone, its authors stating that the full Jones matrix would require depth
multiplexing.
Under Single-shot it means that one spectrometer frame
carries the whole spectrum or depth scan, without polarization
\cite{kalashnikov2016,vanselow2020,zhang2025h} and, in its single-shot envelope mode,
Ref.~\cite{kaufmann2022}; in the Fisher row it
refers to Ref.~\cite{zhang2025h} alone, since Ref.~\cite{panda2026} treats a switched
phase, while the
Fourier-transform and time-domain variants scan a delay line
\cite{lindner2019,mukai2021,paterova2017}; for dual-comb polarimetry it means one
polarizer setting per spectrum in Ref.~\cite{hinrichs2023}, whereas
Ref.~\cite{koresawa2026} reads the full matrix from a single acquisition carrying
three delay-separated bursts. Under
$\lambda$ it means that spectral bins serve calibration \cite{braaf2014} or carry a
polarization encoding spread over the source spectrum \cite{jones2026} and feed a
single matrix; the Fisher row carries no $\lambda$ mark because the parameters
estimated there are scalar per sample. Under FIM it means an optimal-design theory
built on the equally weighted variance rather than on a Fisher matrix
\cite{alenin2014}. Under Depth it marks our own row: the depth scan is available in
the same frame but not demonstrated here. Coincidence
schemes, which detect both photons, are discussed in Sec.~\ref{sec:discussion}. The
combination of all columns in the last row has not been realized before.}
\begin{ruledtabular}
\begin{tabular}{lccccccl}
 & Und. & Single-shot & JM & $\lambda$ & Depth & FIM & Ref. \\
\hline
Process tomography (qudit, sequential)      & \checkmark &              & (\checkmark) &              &              &            & \cite{rajeev2026b,rajeev2026} \\
Polarimetry (multi-frame)                   & \checkmark &              & (\checkmark) &              &              &            & \cite{paterova2018b,chakraborty2025,oglialoro2025,vasikonis2026,arya2025b} \\
Single-frame amplitude and phase            & \checkmark & \checkmark   &              &              &              &            & \cite{haase2022,pearce2024,leontorres2024,topfer2024} \\
Spectroscopy (no polarization)              & \checkmark & (\checkmark) &              & \checkmark   &              &            & \cite{kalashnikov2016,lindner2019,mukai2021,kaufmann2022} \\
OCT (no polarization)                       & \checkmark & (\checkmark) &              &              & \checkmark   &            & \cite{paterova2017,vanselow2020} \\
Fisher analysis (scalar parameters)         & \checkmark & (\checkmark) &              &              &              & \checkmark & \cite{panda2026,zhang2025h} \\
\hline
Channeled spectropolarimetry (classical)    &            & \checkmark   & (\checkmark) & \checkmark   &              & (\checkmark) & \cite{oka1999,hagen2007,alenin2014,dubreuil2007,hagen2021} \\
Dual-comb polarimetry (classical)           &            & (\checkmark) & \checkmark   & \checkmark   &              &            & \cite{koresawa2026,hinrichs2023} \\
Jones-matrix OCT (classical)                &            & \checkmark   & \checkmark   & (\checkmark) & \checkmark   &            & \cite{baumann2012,lim2012,braaf2014,jones2026} \\
\hline
\textbf{This work}                          & \checkmark & \checkmark   & \checkmark   & \checkmark   & (\checkmark) & \checkmark & --- \\
\end{tabular}
\end{ruledtabular}
\end{table*}

\section{Scheme}
\label{sec:scheme}

The scheme rests on three ingredients, each with its own subsection. A folded,
three-arm interferometer places the sample in the idler arm and returns signal and
pump to their own crystals (Sec.~\ref{sec:folded}). A diagonally polarized pump makes
both crystals generate on both passes, which is what turns a measurement of the
idler's polarization \emph{state} into a measurement of the polarization
\emph{process} it undergoes (Sec.~\ref{sec:diagonal}). And the birefringence of the
crossed crystals, together with one delay plate, gives every interfering amplitude
pair its own fringe period, so that all of them can be told apart in a single
spectrometer trace (Sec.~\ref{sec:walkoff}).

\subsection{Folded interferometer and three-arm geometry}
\label{sec:folded}

The source is a pair of identical type-0 nonlinear crystals C1/C2 (same length $L$,
same poling period), crossed with their optic axes at $90^\circ$ and double-passed in
a Michelson configuration
(Fig.~\ref{fig:setup}). Dichroic mirrors behind the crystals separate pump, signal, and
idler into three arms terminated by mirrors. Each end mirror sits in the Fourier plane
of the crystals, so that the returning beams retrace their spatial modes and the pairs
of the two passes are indistinguishable in transverse momentum. The scheme is treated
throughout as single-mode in space; the instrument contrast of Sec.~\ref{sec:recipe}
is the overlap of the probe and reference pairs in that mode, the same quantity that
sets the visibility of an induced-coherence interferometer through the heralding
efficiency of its spatial modes \cite{kim2023}. Spot size, mode number, and the
imaging of extended samples are beyond the present scope. The unknown channel $M$ acts on the idler
in its arm. Throughout, $M$ is the \emph{round-trip} Jones matrix of that arm
(sample, end mirror, sample again),
\begin{equation}
M = \begin{pmatrix} |\Mhh|\,e^{i\arg\Mhh} & |\Mhv|\,e^{i\arg\Mhv} \\[2pt]
|\Mvh|\,e^{i\arg\Mvh} & |\Mvv|\,e^{i\arg\Mvv} \end{pmatrix},
\label{eq:jones}
\end{equation}
where $M_{ij} = \langle i | M | j \rangle$, the first index the output and the second
the input polarization, so that
$\Mvh$ maps an $H$ idler onto $V$: four moduli and, since a global phase is
unobservable, three relative phases, seven real parameters. For a reciprocal sample
it is symmetric, so only five of the seven are independent. Its antisymmetric part is then a
background-free signature of non-reciprocity (Sec.~\ref{sec:discussion}).
The other two arms carry fixed optics only. The signal arm holds a birefringent
delay plate $\mathcal{P}(T)$, whose single-pass group delay $T$ between the two
plate axes sets the carrier splitting $d_1$ of
Sec.~\ref{sec:walkoff}, and a quarter-wave plate, which in double pass acts as a
half-wave plate. We write its Jones matrix as the reflection $R(A)$ of
Sec.~\ref{sec:rates}, with $A$ twice the axis angle. At the working point the
axis sits at $22.5^\circ$, so $A = 45^\circ$. The pump arm is
terminated by a bare end mirror. At the balanced working point of
Sec.~\ref{sec:working} it needs no wave plate. After the second
pass a polarizing beam splitter (PBS) splits the returning signal into its $H$
and $V$ components. The spectra of its two output ports $D_H$ and $D_V$ are
recorded in the same frame on two spectrometers (Fig.~\ref{fig:setup}). Recording the ports
in the same frame matters, because one of the three Jones phases is a difference
between a $D_H$ and a $D_V$ family (Sec.~\ref{sec:recipe}); the two spectrometers
must therefore share one exposure window, a hardware requirement rather than a
calibration. All photon-budget and read-noise statements below count per frame,
i.e.,\ per pair of traces.

Compared to the two-source Zou--Wang--Mandel geometry
\cite{zou1991,fuenzalida2022b}, in which only the idler paths are aligned and the two
signals meet on a beam splitter, the fold returns signal \emph{and} idler of the first
pass into the crystals, which is why the signal arm carries a wave plate at all
(Sec.~\ref{sec:diagonal}). Beyond that, the fold buys three things. First, pump-amplitude matching
between the first (probe) and the second (reference) pass is automatic, because one
and the same pump beam drives both passes. In the two-source geometry of
Ref.~\cite{fuenzalida2022b} the reference source is instead pumped with $\sqrt2$
times the amplitude of the probe source ($b_2 = \sqrt2\,b_1$ in its Supplemental
Material). That compensates the equal split of the reference pairs between $H$
and $V$. Second, the two polarization channels are read out in
parallel on the two PBS ports. Third, all arm phases are common mode. Three things
the fold does not make common mode: the wavelength registration of the two
spectrometers, a calibration quantity that Sec.~\ref{sec:chirp} budgets and that is
tracked within every frame; the crystal temperature, which moves the calibration
phases $\eta_f$ (Sec.~\ref{sec:feasibility}); and the channel used in the calibration
frame (Sec.~\ref{sec:recipe}). The last two sit in that frame, whose repetition rate
they set.
A fixed signal--idler group-delay imbalance $\dT$ maps the
interferometer phase onto a spectral carrier, $\Phi(\Omega) = \Phz + \Omega\,\dT$,
so that a spectrometer frame can replace the phase scan \cite{vanselow2020,kaufmann2022}. Here $\Omega$ denotes the
detuning of the signal from its phase-matched center frequency $\omega_{s0}$.
Because the pump is monochromatic, energy conservation places the partner idler
at $\omega_{i0} - \Omega$, so one variable labels both photons
(App.~\ref{app:rates}).

\subsection{Diagonal pump: from state to coherent process tomography}
\label{sec:diagonal}

The pump angle $\chi$ decides what is measured. We start at $\chi = 0$, where the
scheme reduces to state tomography, and then increase it.
With a horizontally polarized pump ($\chi=0$) only C2 generates on the first pass, and
the scheme measures the polarization \emph{state} of the returning idler, the folded
version of Ref.~\cite{fuenzalida2022b}. That state is a $2\times2$ density matrix
$\rho_I$ with $H$ population $\Pi_H$, off-diagonal element
$\rho_{HV} = \purity \sqrt{\Pi_H(1-\Pi_H)}\, e^{-i\xi}$, and hence three parameters: $\Pi_H$,
the relative phase $\xi$ between $H$ and $V$, and the coherence $\purity$, with
$0 \le \purity \le 1$ the ratio of the actual to the maximal coherence. In this limit the pump is returned
polarized at the angle $B = A$, where $B$ is the return-pump angle, the polarization
angle of the pump on its second pass through the crystals (Sec.~\ref{sec:rates}).
For $\chi = 0$ that \emph{does} require a pump-arm wave plate, the flip-in plate of
Fig.~\ref{fig:setup}, unlike at the process working point (Sec.~\ref{sec:working}). With the spontaneous parametric
down-conversion (SPDC) spectrum $S(\Omega)$ the two port spectra, i.e.,\ the count
rate per spectrometer pixel $I_P(\Omega)$ at PBS port $P$, are (App.~\ref{app:rates})
\begin{equation}
I_P(\Omega) \propto S(\Omega)\, \mathcal{W}_P \big[1 + \mathcal{V}_P
\cos(\Phz + \Omega\,\Delta_P + \xi_P)\big],
\label{eq:staterates}
\end{equation}
\begin{equation}
\mathcal{V}_H = t_H \sqrt{\Pi_H},
\qquad
\mathcal{V}_V = t_V\, \purity \sqrt{1 - \Pi_H},
\label{eq:statevis}
\end{equation}
with $P \in \{H, V\}$, port weights $\mathcal{W}_H = \cos^2\!A$, $\mathcal{W}_V = \sin^2\!A$, distinct
carriers $\Delta_H \neq \Delta_V$, fringe phases $\xi_H = 0$ (phase anchor) and
$\xi_V = \xi$, and calibrated transmission amplitudes $t_{H,V}$ (those of
Ref.~\cite{fuenzalida2022b}). Two visibilities and one relative fringe phase determine
the state. Because the fold matches probe and reference amplitudes automatically, both
visibilities are independent of the signal wave plate angle $A$: that plate is a
pure photon-budget knob between the two PBS ports. Pumping at
angle $\chi$ instead entangles the first-pass pair,
\begin{equation}
\ket{\psi_1} = \cos\chi \, \ket{H_s H_i} + e^{i\kappa}\sin\chi\,\ket{V_s V_i},
\label{eq:pass1}
\end{equation}
so the idler enters the unknown channel in both basis states: all four elements
of the Jones matrix $M$ are illuminated in a single frame. The double-passed signal
wave plate $A$ does more than split flux here. Without it the signal
polarization would be a which-source tag that forbids interference of the off-diagonal
amplitudes, and only $\Mhh$, $\Mvv$ would survive. The wave plate \emph{erases the
which-source information} and thereby unlocks $\Mhv$, $\Mvh$. Its angle distributes
the photon budget between diagonal ($\cos A$) and off-diagonal ($\sin A$) families.

\subsection{Birefringent walk-off as a frequency multiplexer}
\label{sec:walkoff}

Were all interfering amplitude pairs to share the same interferometer phase $\Phi$,
each spectrometer trace would collapse to a single sinusoid: three observables per
frame. Every further parameter would then demand a new setting, the route of
Ref.~\cite{rajeev2026}. The birefringence of the crossed crystals breaks this
degeneracy. The interfering families (first-pass $H$/$V$ pair $\times$ $H$/$V$
reference) have different group-delay budgets (generation in C2 versus C1, and
$H$ versus $V$ transits) and hence different spectral fringe periods. Where the natural
walk-off is insufficient it can be engineered. A birefringent delay plate in the
signal arm sets the splitting $d_1$, which the crystal walk-off renormalizes
by a known offset: $13\%$ for our crystals (App.~\ref{app:transits}). Once the period differences amount to a few fringes
across the bandwidth, the families are individually resolvable in a single spectrum.
This is frequency multiplexing --- the mechanism behind heterodyne channel
separation and behind passive polarization-delay Jones-matrix OCT
\cite{baumann2012,lim2012,ju2013}. Its closest classical relative is
channeled spectropolarimetry, where thick birefringent retarders place the Stokes
parameters of a beam, or with a modulated generator the Mueller elements of a sample
\cite{hagen2007}, on distinct spectral carriers, all recovered from one Fourier
transform of one spectrum \cite{oka1999}. That technique comes complete
with its own optimal-design theory \cite{alenin2014} and has been demonstrated
experimentally \cite{dubreuil2007,hagen2021}. Classical single-frame
Jones-matrix measurement is also reached holographically, by common-path
polarization holography under partially coherent illumination
\cite{liuhaocun2025}.

\section{Detection rates and fringe families}
\label{sec:rates}

This section derives the measurement equations, the one structure on which the
readout, the design, and the information analysis of the later sections are built.
Both spectrometer traces share one structure: a channel-independent offset plus one
cosine per interfering amplitude pair. We parametrize the double-passed signal wave plate
as the reflection
$R(A)\!:\ \ket{H}\to\cos A\ket{H}+\sin A\ket{V},\ \ket{V}\to\sin A\ket{H}-\cos A\ket{V}$
(the plate axis sits at $A/2$, Sec.~\ref{sec:folded}),
and the returning pump as $(b_H,\, b_V e^{i\theta})$ with $b_H=\cos B$, $b_V = \sin B$.
The unknown channel is the Jones matrix $M$ of Eq.~(\ref{eq:jones}), with losses
included via the vacuum-ancilla
formalism of Ref.~\cite{rajeev2026}. With the SPDC spectrum $S(\Omega)$ of
Sec.~\ref{sec:scheme}, both traces take the master form
\begin{equation}
I_P(\Omega) \propto S(\Omega)\Big[\mathcal{B}_P + \!\!\sum_{f\in\mathcal F_P}\!\! 2\, w_f |X_f|
\cos\!\big(\phi_f(\Omega)\big)\Big],
\label{eq:master}
\end{equation}
\begin{equation}
\phi_f(\Omega) = s_f\,\Phz + \Omega\,\Delta_f + \arg X_f + \eta_f ,
\qquad s_f\in\{0,1\},
\label{eq:phases}
\end{equation}
where $P\in\{H,V\}$ labels the PBS port, $\mathcal{B}_P$ is the non-interfering
offset of that port, and the sum runs over the set $\mathcal F_P$ of fringe
families it carries (Table~\ref{tab:families}). For each family, $w_f$ is a known
setting-dependent amplitude weight, unlike the intensity weight $\mathcal{W}_P$ of
Eq.~(\ref{eq:staterates}); $X_f$ is the carried channel combination, $\Delta_f$ the
carrier delay, and $\eta_f$ a calibration phase (birefringent crystal phases and
delay-plate offsets). The flag $s_f$ says whether the family carries the global
interferometer phase $\Phz$. It does ($s_f = 1$) whenever a first-pass amplitude
interferes with a second-pass amplitude, because $\Phz$ is the phase the second
pass acquires in the three arms relative to the first. The two bright families
per port are of this kind. The fifth carrier interferes two first-pass amplitudes
with each other, so $\Phz$ cancels and $s_f = 0$ (Table~\ref{tab:families}).
Three phase symbols run through the paper: $\Phi$ for interferometer phases,
$\phi_f$ for the total phase of a family, and $\varphi$ for phases acquired in the
crystals and plates, carrying a superscript where it belongs to a single transit.

The carried combination $X_f$ is always the same kind of object. Only the signal is
detected, so two signal amplitudes interfere with a contrast given by the overlap of
the idler states that accompany them. This is the rule at the root of every
measurement with undetected photons, from Zou, Wang, and Mandel onwards
\cite{zou1991,wangzoumandel1991}, and $X_f$ is that overlap. For F1--F4 one idler was born on the first pass in the state
$\ket{i_{Cj}}$ of its crystal ($\ket{i_{C2}} = \ket{H}$, $\ket{i_{C1}} = \ket{V}$)
and made the round trip through the channel, while the partner idler was born on the
second pass in $\ket{i_{Ci}}$ and never saw the channel. Their overlap is
$\langle i_{Ci}|M|i_{Cj}\rangle = M_{ij}$, which is how the Jones matrix enters
Table~\ref{tab:families} entry by entry. For F5/F6 both idlers were born on the first
pass, one in each crystal, and both made the round trip; their overlap is the subject of
Sec.~\ref{sec:witness}. The offsets are
\begin{equation}
\mathcal{B}_H = \cos^2\!\chi\cos^2\!A + \sin^2\!\chi\sin^2\!A + b_H^2,
\label{eq:bgH}
\end{equation}
\begin{equation}
\mathcal{B}_V = \cos^2\!\chi\sin^2\!A + \sin^2\!\chi\cos^2\!A + b_V^2,
\label{eq:bgV}
\end{equation}
Their sum is $\mathcal{B}_H + \mathcal{B}_V = 2$ for every $\chi$, $A$, and $B$,
because the first and the second pass each contribute one unit of pair flux
whatever the settings. The settings therefore only redistribute a fixed number of
detected photons between the two ports and among the fringe families. This is
what makes the Fisher comparison of Sec.~\ref{sec:fim} fair by construction: every
working point is compared at the same photon number.

Table~\ref{tab:families} lists the fringe families. F1--F4 pair a first-pass (probe)
amplitude with a reference amplitude and carry one Jones element each on its own
carrier. The geometry supplies a fifth carrier at $d_1$, one per port: the
\emph{self-interference} of the two first-pass components. Both of its amplitudes
belong to pass one, so $\Phz$ drops out ($s_f = 0$). For
$\dT \gtrsim 2 d_1$ the carrier sits far from F1--F4: its spacing from the nearest
bright carrier is $\dT - d_1$, many peak widths. It is moreover dark for every
physical channel, the \emph{dark carrier at $d_1$}, which is what lets it
serve as a null test (Sec.~\ref{sec:witness}).

\begin{table}[t]
\caption{\label{tab:families}%
The fringe families of the diagonally pumped folded interferometer,
Eqs.~(\ref{eq:master})--(\ref{eq:phases}). The columns are: the family $f$, the PBS
port on which it appears, its weight $w_f$, the channel combination $X_f$ it carries,
its carrier delay $\Delta_f$, and whether it carries the interferometer phase $\Phz$
($s_f = 1$). F1--F4 carry the four Jones elements. F5/F6 form the dark carrier
at $d_1$: it is $\Phz$-free, and its amplitude is the overlap of the two first-pass
idlers born in C1 and C2, which vanishes for every physical channel
(Sec.~\ref{sec:witness}). The carriers form an exact equidistant ladder, counting
how often each family passes the delay plate, zero to twice. Here $d_1$ is the plate delay
$T$ renormalized by the crystal walk-off, and the whole ladder is quoted up to a
common origin shift (both in App.~\ref{app:rates}). F2 and F3 coincide in carrier
and are told apart by the port alone. The pump return phase $\theta$ and the pair
phase $\kappa$ of Eq.~(\ref{eq:pass1}) are calibration constants; they enter only as
the sum $\theta + \kappa$, booked in $\eta_3$ (App.~\ref{app:rates}), not in $X_f$.}
\begin{ruledtabular}
\footnotesize
\begin{tabular}{llllll}
$f$ & Port & Weight $w_f$ & Carries $X_f$ & $\Delta_f$ & $\Phz$? \\
\hline
F1 & $D_H$ & $b_H\cos\chi\cos A$            & $\Mhh$              & $\dT$        & yes \\
F2 & $D_H$ & $b_H\sin\chi\sin A$            & $\Mhv$              & $\dT + d_1$  & yes \\
F3 & $D_V$ & $b_V\cos\chi\sin A$            & $\Mvh$              & $\dT + d_1$  & yes \\
F4 & $D_V$ & $-b_V\sin\chi\cos A$           & $\Mvv$              & $\dT + 2d_1$ & yes \\
F5 & $D_H$ & $\cos\chi\sin\chi\cos A\sin A$ & $\langle i_{C1}|i_{C2}\rangle = 0$ & $d_1$ & no \\
F6 & $D_V$ & $-\cos\chi\sin\chi\sin A\cos A$& $\langle i_{C1}|i_{C2}\rangle = 0$ & $d_1$ & no \\
\end{tabular}
\end{ruledtabular}
\end{table}

\section{Single-frame readout}
\label{sec:readout}

Section~\ref{sec:rates} says which cosines a spectrometer trace contains. This
section is about telling them apart in a single frame, and the point of view changes
accordingly, from photon physics to signal processing. The tool is the Fourier
transform of each trace along its frequency axis: every family becomes a peak at its
own delay, and the six families sort themselves onto a ladder of delays. We first
give the recipe (Sec.~\ref{sec:recipe}), then ask how far apart the peaks must sit
and what fixes their positions (Sec.~\ref{sec:placement}), explain why one carrier
stays dark for every channel and how that serves as a built-in null test
(Sec.~\ref{sec:witness}), run the whole chain on a simulated frame
(Sec.~\ref{sec:demo}), and finally extend the readout to samples whose response
varies across the spectrum (Sec.~\ref{sec:budget}).

\subsection{Delay-space picture and recipe}
\label{sec:recipe}

A spectrometer trace is not one number but ${\sim}10^3$ pixels across the bandwidth.
Each family is a cosine of known period riding on it. Taking the Fourier transform of a
trace with respect to its frequency axis, the detuning $\Omega$, turns each cosine into a
peak at its delay $\Delta_f$: the transform shows one peak per family in delay space,
exactly as an OCT A-scan shows one peak per interfering pair at its group-delay difference. Peak
height gives the family amplitude, complex peak phase the family phase. Extracting all
family parameters thus amounts to one fast Fourier transform (FFT) followed by a
read-off of the two bright peaks per
port; no iterative fit is needed. The readout recipe is:
\begin{enumerate}
\setlength{\itemsep}{2pt}\setlength{\parsep}{0pt}\setlength{\parskip}{0pt}
\item Fit each trace as a sum of sinusoids at the four ladder carriers
$\{\dT,\, \dT{+}d_1,\, \dT{+}2d_1,\, d_1\}$, known from design or from a
reference frame, each carrier taken with its calibrated chirp, the slow drift of the
carrier delay across the window imposed by crystal dispersion
(Sec.~\ref{sec:chirp}). This is a per-port least-squares fit at the known carriers.
Each port shows two bright
peaks: port $D_H$ carries $\dT$ and $\dT{+}d_1$, port $D_V$ carries $\dT{+}d_1$
and $\dT{+}2d_1$. The fourth carrier, $d_1$, is present on both ports but dark by
design; it is monitored as a null test (Sec.~\ref{sec:witness}).
\item \emph{Moduli}: $|\Mhh|, |\Mhv|, |\Mvh|, |\Mvv|$ are the family amplitudes divided by
the known weights $w_f$ and the common instrument contrast.
\item \emph{Phases}: F1--F4 share $\Phz$, leaving three $\Phz$-free differences:
$\arg\Mhv - \arg\Mhh$ (F2$-$F1), $\arg\Mvv - \arg\Mvh$ (F4$-$F3), and
$\arg\Mvh - \arg\Mhh$ (F3$-$F1), each up to its calibration constant. These three ladder differences
determine the phase basis
$(\arg\Mhv{-}\arg\Mhh,\allowbreak\ \arg\Mvh{-}\arg\Mhh,\allowbreak\
\arg\Mvv{-}\arg\Mhh)$ of Sec.~\ref{sec:fim} invertibly. Together with the four moduli of
step 2, these three phases make seven real numbers, which is exactly a Jones matrix
modulo its global phase: the reconstruction is complete. The only instrument quantities
entering the differences are the two calibration combinations $\varepsilon$ and
$\theta{+}\kappa$ of App.~\ref{app:rates}, fixed in step 5; the third constant,
$\eta_1$, drops out of the differences.
\item \emph{Consistency}: in the $\chi = 0$ state limit only F1
($\Mhh \to t_H\sqrt{\Pi_H}$) and F3
($\Mvh \to \purity\, t_V \sqrt{1-\Pi_H}\, e^{+i\xi}$; the conjugate index order of
$\rho_{HV}$ accounts for the sign) survive, reproducing the state
visibilities of Eq.~(\ref{eq:statevis}): the state tomography of
Ref.~\cite{fuenzalida2022b} is recovered as a special case of the process rates.
\item \emph{Calibration}: one reference frame with a \emph{known} channel that lights
up all four families determines the two calibration combinations of
App.~\ref{app:rates} that enter the phase differences ($\varepsilon$ and
$\theta{+}\kappa$), the carriers,
and the chirp of each carrier, all from that single reference frame. The third
constant of App.~\ref{app:rates}, $\eta_1$, is a pure absolute anchor and is never
needed for the reconstruction. A known
wave plate in the idler arm is one such
channel; Sec.~\ref{sec:chirp} shows that for a reciprocal object the constants can
also be read from measurement frames. The empty arm, $M = \openone$, will not do on
its own, because it leaves F2/F3 dark; together with one measurement frame of any
reciprocal object, whose retardance and axis need not be known, it fixes both
constants without a \emph{calibrated} object (Sec.~\ref{sec:chirp}).
The $\eta_f$ drift with crystal temperature, which sets how often this frame has to
be repeated (Sec.~\ref{sec:feasibility}).
\end{enumerate}

\subsection{Carrier placement and separability}
\label{sec:placement}

Separation is a per-trace condition: each port carries two bright families, spaced
by $d_1$ on the ladder, plus the dark one ($D_H$: F1, F2, F5; $D_V$: F3, F4, F6).
The splitting $d_1$ must be nonzero and large enough that neighboring peaks are
resolved (peak width $\approx 1/\Delta\nu$). All carriers must lie inside the window
$1/\Delta\nu \ll \Delta_f \ll 1/\delta\nu$ set by the bandwidth $\Delta\nu$ and
the spectrometer resolution $\delta\nu$; below the resolution is quoted in
wavelength, $\delta\lambda = \lambda^2\,\delta\nu/c$.
At delay zero sits the zero-delay (DC) hump, the transform of the non-interfering
term $\mathcal{B}_P\,S(\Omega)$, whose width in delay is set by the smoothness of
the spectrum. As in off-axis holography, every carrier must clear this zero-order hump.
The binding case is the lowest carrier, the dark one at $d_1$, since the bright ones
sit at $\dT \gtrsim 2 d_1$ and beyond.
The ladder delays are group delays taken at the band center. The crystals are
dispersive, so the delay a family actually carries changes slowly across the window,
and it changes by a different amount for each family. What counts is the difference
in crystal transits between the two amplitudes that interfere, the probe pair and its
reference pair. This difference grows with every step $d_1$ up the ladder, by one
transit in the non-generating ($y$) polarization, whose mid-IR idler carries almost
all of the dispersion, plus one pass of the delay plate (App.~\ref{app:transits}). Hence F4 at $\dT{+}2d_1$
drifts about twice as much as F2 and F3 at $\dT{+}d_1$, and F1 at $\dT$ hardly at
all. Toward the band edges the two
bright peaks of a port therefore move closer together than their nominal spacing
$d_1$. The gap between them has to stay wide enough to absorb this drift. The
spectrum carries little weight at the edges, and the demodulation is not affected,
because it uses the calibrated dispersion (Sec.~\ref{sec:chirp}).
Note that the
ladder makes F2 and F3 exactly degenerate: they are
separated by the PBS alone. Finite polarization extinction therefore maps each onto the
other's genuine peak rather than onto empty delay space. This is quantified in
App.~\ref{app:transits}, which also shows that symmetry-breaking alternatives are
impractical and gives a deterministic subtraction recipe.

\subsection{The dark carrier at \texorpdfstring{$d_1$}{d1}: a built-in null test}
\label{sec:witness}

The dark carrier at $d_1$ deserves care, because a naive calculation gets it
wrong. Inserting a bare Jones matrix $M$ into the rates predicts an F5 amplitude
proportional to the column coherence $\dephw$. The physical picture is simpler.
The $d_1$ carrier interferes the two first-pass amplitudes: an idler born in C2, in
polarization state $\ket{i_{C2}}$, and an idler born in C1, in state
$\ket{i_{C1}}$. For ideally crossed crystals these are $\ket{H}$ and $\ket{V}$.
Both idlers pass through the channel, and the interference term is weighted by the
overlap of the two \emph{output} states. Now, a lossy or decohering channel is not
unitary on the idler alone, but it becomes unitary once we include its
\emph{environment}: the modes into which the channel absorbs or scatters light and
which are never detected (the ancilla modes of App.~\ref{app:rates}). Writing the
whole action as a unitary $U$ on idler plus environment, both output states are
$\ket{\psi^{\rm out}_{Ck}} = U\,\ket{i_{Ck}}\ket{{\rm env}_0}$ with the same
initial environment state. A unitary preserves overlaps,
$\langle b|U^\dagger U|a\rangle = \langle b|a\rangle$, hence
\begin{equation}
\langle \psi_{C1}^{\rm out}|\psi_{C2}^{\rm out}\rangle = \langle i_{C1}|i_{C2}\rangle
\label{eq:darkoverlap}
\end{equation}
for every channel, coherent or not. This is the entry $X_5 = X_6 =
\langle i_{C1}|i_{C2}\rangle$ of Table~\ref{tab:families}: the F5 amplitude equals the
\emph{input} overlap of the two idlers, and for crossed crystals that overlap is zero.
Where does the naive
$\dephw$ come from? It is the overlap of the \emph{transmitted} parts alone. The parts
the channel absorbs or scatters enter the environment with amplitudes $C$, and
$M^\dagger M + C^\dagger C = \openone$ forces their overlap to be exactly $-\dephw$
(App.~\ref{app:rates}). A $45^\circ$ polarizer as the channel makes this vivid: both
transmitted idlers emerge diagonally polarized (overlap $+\tfrac12$), but the absorbed
antidiagonal components enter the same environment mode with opposite amplitude signs
(overlap $-\tfrac12$). The books always balance. In the Heisenberg picture the same
statement is immediate: the two probe amplitudes multiply the orthogonal idler
\emph{input} modes born in C2 and C1, and the channel, which acts only afterwards,
never enters those coefficients at all. Both
derivations, and an explicit numerical dilation model, agree: F1--F4 match
Table~\ref{tab:families} while the $d_1$ carrier vanishes for arbitrary channels
(App.~\ref{app:rates}). This is consistent with visibility-only readouts, in which a
unitary rotation of the idler acts as a which-path marker and lowers the fringe
contrast \cite{paterova2018b,chakraborty2025}: there the rotation changes the one
overlap the single fringe carries, while here it redistributes amplitude among
F1--F4 and leaves the input overlap of the dark carrier untouched.

Far from being wasted, the dark carrier does three jobs.
(i)~Its amplitude equals the input overlap $\langle i_{C1}|i_{C2}\rangle$ of
Eq.~(\ref{eq:darkoverlap}) and is therefore independent of the channel. Any light at
$d_1$ thus measures source or alignment error. If the crystals are crossed at
$90^\circ + \epsilon_\times$ instead of $90^\circ$, the overlap is $\approx
-\epsilon_\times$ times the spectral overlap of the two generation processes, so the
dark carrier lights up in \emph{first} order in the crossing error; crystal cross-talk
adds to it. A common rotation of both crystals against the PBS basis leaves the overlap
zero, only the relative deviation from $90^\circ$ counts. This is the tool that
verifies the crossed-crystal alignment itself, per frame and independently of what sits
in the idler arm.
(ii)~It flags readout violations that mix carriers: camera nonlinearity or
layered-sample autocorrelation terms (Sec.~\ref{sec:discussion}). One violation it
cannot see is the PBS-extinction ghost, which lands exactly on a genuine peak
rather than at $d_1$ (App.~\ref{app:transits}).
(iii)~Being $\Phz$-free and photon-neutral, it runs continuously while F1--F4 carry
the measurement.

\subsection{Numerical demonstration}
\label{sec:demo}

\begin{figure*}[t!]
\includegraphics[width=0.96\textwidth]{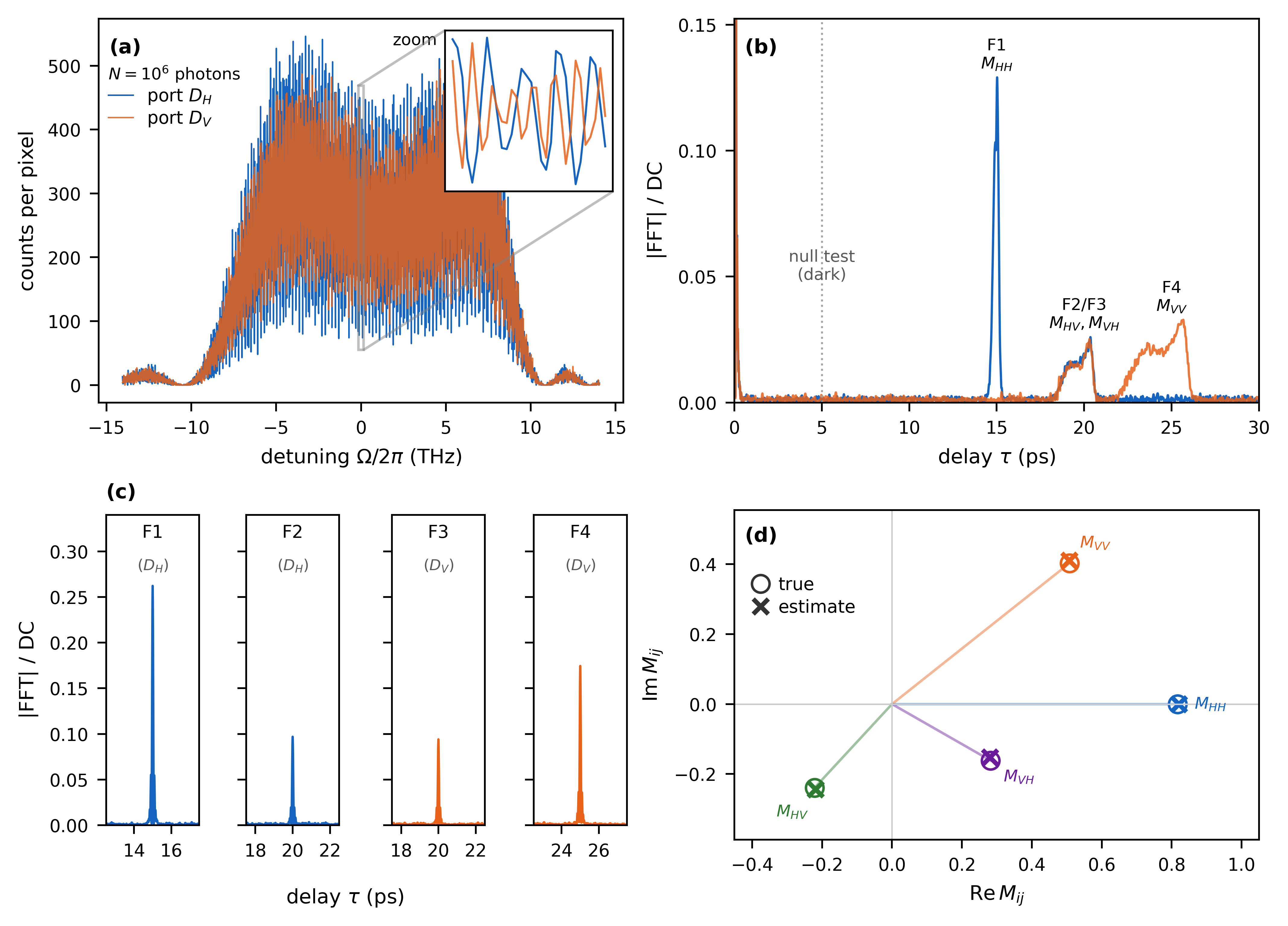}
\caption{\label{fig:demo}%
Single-frame readout, simulated end to end with the crystal dispersion of
Sec.~\ref{sec:chirp} included. The carriers sit at the nominal placement
$\dT = 15$\,ps and $d_1 = 5$\,ps, net of the crystal walk-off
(App.~\ref{app:transits}). The channel is a generic non-reciprocal contraction
(retarder $0.8$\,rad at $30^\circ$, diattenuation $0.9/0.7$, circular rotation
$20^\circ$), so all four Jones elements are populated and $\Mhv \neq \Mvh$.
(a)~One simulated frame: both spectrometer traces, pixelated at the $20$\,pm pitch
of Table~\ref{tab:numbers} with a Gaussian instrument function of the same width and
per-pixel Poisson noise, $N = 10^6$ detected photons per frame over both traces. The
spectral envelope is the phase-matching envelope of App.~\ref{app:sellmeier}, side
lobes included. The inset resolves the superposition of the three fringe periods.
(b)~Plain Fourier transform of each trace, whose delay axis $\tau$ carries the
carriers $\Delta_f$; the DC hump at zero delay is clipped,
the noise floor lies at $1\times10^{-3}$.
The families sit on the ladder: F1 at $\dT$ on $D_H$, F2 on $D_H$ and F3 on $D_V$
coinciding at $\dT{+}d_1$ and separated by the port alone, F4 at $\dT{+}2d_1$
on $D_V$. The dark carrier at $d_1$ shows no peak. Only F1 is a sharp peak; the
chirped families F2--F4 appear as humps of $1.5$--$3$\,ps width, F4 visibly shifted
from its nominal position.
(c)~Matched demodulation inside the four $\pm2.5$\,ps windows of
Sec.~\ref{sec:budget}: each trace is transformed with the calibrated chirp of the
family targeted in that window. Every family collapses to the transform limit of
the phase-matching envelope ($0.053$\,ps, slightly narrower than the $1/\Delta\nu$
delay bin of Table~\ref{tab:numbers}) at its nominal delay. The peaks rise because the energy smeared
over the humps of (b) is concentrated into the transform limit; these heights are
what the estimator reads as amplitudes.
(d)~Jones matrix read from this one frame by the matched estimator (crosses)
against the ground truth (circles).}
\end{figure*}

Figure~\ref{fig:demo} shows the recipe run end to end on one simulated frame, for a
deliberately generic channel, neither reciprocal nor unitary, so that no matrix
element is trivially zero. Throughout we quote two photon budgets: $N = 10^6$
detected photons, summed over both traces, for a single frame, and $N = 10^8$ for a
campaign of $100$ such frames. With the single-frame budget and pure Poisson pixel
noise, the two traces of panel~(a) transform into the delay spectra of panel~(b).
The plain transform already shows the ladder, but with the crystal dispersion of
Sec.~\ref{sec:chirp} in the model only F1 is transform limited; the chirped families
are humps. The matched demodulation of panel~(c), which uses the calibrated chirp of
the targeted family, restores the sharp peak at its nominal delay, and reading off
the complex amplitudes returns all
seven parameters with errors below $10^{-2}$ [panel~(d)]; the largest deviation is
$1.8$ standard errors of the bound of Sec.~\ref{sec:fim}. The reconstruction includes
the non-reciprocal asymmetry between $\Mhv$ and $\Mvh$. For a reciprocal channel the
same estimator returns $\Mhv = \Mvh$ within error \emph{without assuming
reciprocity}, a free consistency check on top of the null test.

\subsection{Spectrally resolved channels and the time--bandwidth budget}
\label{sec:budget}

If $M$ depends on wavelength, each family becomes a modulated carrier (fast fringe at
$\Delta_f$, slow envelope $M_f(\Omega)$). It is demodulated exactly as in AM radio or
off-axis holography: window each delay-space peak, transform back, and obtain the full
complex envelope $M_f(\lambda)$ per family, still within the one frame. The three
$\Phz$-free differences of Sec.~\ref{sec:recipe} then live on as complex ratios of
demodulated envelopes ($\Phz$ cancels pointwise in $\Omega$), so the
wavelength-resolved Jones matrix retains the single-frame property. The
condition is
non-overlap of the delay-space peaks: the internal group-delay spread of the sample
must stay below the carrier spacing $d_1$ (OCT logic: specimen structure = delay
content). The currency is the time--bandwidth product
$N_{\rm slot} = \Delta\nu/\delta\nu$, the number of independent spectral samples per
trace. Because the trace is real valued, its Fourier transform is Hermitian, and only the
positive half of delay space carries independent information: $N_{\rm slot}/2$ delay
bins of width $1/\Delta\nu$ up to the Nyquist ceiling $1/(2\delta\nu)$. Into this half
must fit the DC hump at zero delay and three carriers per trace: the two bright
families and the dark carrier at $d_1$, whose neighborhood must stay empty for the
null test to remain readable. Between the carriers lie guard bands, stretches of delay
space kept free so that the demodulation windows do not overlap once the sample's
group-delay structure broadens the peaks. A demodulation window of $W_{\rm dem}$
delay bins returns $W_{\rm dem}$ independent spectral points, so $N_\lambda$ is
simply the window width in bins. Tiling the positive half with three full windows and the half-window of the DC
hump, with no guard bands left, gives the ceiling $N_\lambda \lesssim N_{\rm slot}/7$.
The conservative placement of Table~\ref{tab:numbers}, with its $\pm2.5$\,ps
demodulation windows, uses about a third of it.
The demodulated envelope is returned on the pixel grid, but windowing is a
low-pass: adjacent samples are correlated, and error statistics must be based on the
$N_\lambda$ independent points, not the pixels. Sharp spectral features
(long delay tails) make the families bleed into each other. The escapes are wider
carrier spacing (coarser $M(\lambda)$), sequential operation with one setting at a
time as in Ref.~\cite{rajeev2026}, applied here bin by bin, or the state limit, in
which each trace carries a single family and full spectrometer resolution is free. Note what extra
spectrometer resolution or bandwidth buys. The spectral resolution of $M(\lambda)$ is
$1/W_{\rm dem}$, set by the demodulation window and hence by the carrier spacing, not by the pixel
pitch. Finer spectrometer resolution $\delta\nu$ widens the Nyquist range, which allows
the carriers to be spaced further apart and the windows to be widened: a finer
$M(\lambda)$. More bandwidth $\Delta\nu$ adds spectral points at the same resolution,
covering a wider range. Neither buys precision per point: the Fisher information per
parameter remains limited by the photon budget and camera noise (Sec.~\ref{sec:fim}),
and spreading a fixed budget over more spectral points lowers the precision of each.

\section{Design and feasibility}
\label{sec:feasibility}

The scheme as presented is parameter-free. The rate structure, the carrier ladder
and the readout follow from the interferometer topology alone. The
information-theoretic comparisons of Sec.~\ref{sec:fim} involve only ratios
which, for pure Poisson noise, do not depend on the photon budget. Camera read
noise is the one exception, flagged as such where it matters. This section anchors the
scheme in one concrete set of numbers: the crossed-crystal ppKTP source
available in our laboratory. Nothing in this
choice is fundamental: any other $\chi^{(2)}$ platform maps onto the same design
equations, and Table~\ref{tab:numbers} doubles as the checklist of source parameters
one would have to recompute for it (walk-off, chirp, temperature drift, spectrometer
resolution, delay-window ceiling). The numbers serve two ends. They show that every
requirement of the scheme (delay plate, oven stability, spectrometer resolution,
delay range) is met by off-the-shelf components, and they fix the operating point,
in particular the carrier placement and the plate delay, of the experiment we plan.

\subsection{Birefringence bookkeeping and compensation}

The crossed pair is birefringent; the transit bookkeeping is done in
App.~\ref{app:rates}, and its upshot is simple. The return trip through both crystals
is common to both probe amplitudes, the pairs generated on the first pass (each photon
crosses one crystal along the poled axis $z$ and the other along $y$), so it creates
no asymmetry. What remains are single transits: the $V$ probe, generated in C1 on the first pass,
crosses C2 on its way to the arms (absorbed into the delay-plate specification,
App.~\ref{app:transits}), and the $H$ reference, generated in C2 on the second pass,
crosses C1 on its way to the detector, which leaves between the $HH$ and $VV$
references the walk-off and the phase familiar from crossed-crystal state tomography,
\begin{equation}
\delta = \frac{1}{c}\big[n_g^{y}(\omega_{s0}) - n_g^{y}(\omega_{i0})\big],
\label{eq:delta}
\end{equation}
\begin{equation}
\varphi_{\rm bir} = L\big[k_y(\omega_{s0}) + k_y(\omega_{i0}) - k_y(\omega_p)\big].
\label{eq:phibir}
\end{equation}
Here $n_g^{y}(\omega)$ is the group index along $y$ and $k_\mu(\omega)$ the
wavenumber in polarization $\mu$. The walk-off $\delta$ is a rate per unit
length; the total reference walk-off over one crystal is $\delta L = -0.77$\,ps
for our parameters (App.~\ref{app:sellmeier}).
Their nonlinear parts are calibrated with the chirp (Sec.~\ref{sec:chirp}); the
constant of $\varphi_{\rm bir}$ comes from the reference frame and drifts with crystal
temperature by $-16$\,mrad/(K\,mm) (App.~\ref{app:sellmeier}), so
for $L = 2.55$\,mm an oven holding $25$\,mK/mrad of tolerated drift suffices.
Which crystal faces the arms matters only in the state limit $\chi \to 0$: generating
the $H$ pair in the arm-facing crystal keeps its probe and reference within one
transit of each other (App.~\ref{app:rates}). At $\chi = 45^\circ$ the exchange is a
relabeling of $H$ and $V$.

\subsection{Laboratory numbers}

\begin{table}[htb]
\caption{\label{tab:numbers}%
Design numbers for the ppKTP implementation. The Sellmeier numerics are
collected in App.~\ref{app:sellmeier}; the carrier placement is the working example
used throughout. QPM = quasi-phase-matching, FWHM = full width at half maximum,
DGD = differential group delay.}
\makeatletter
\let\revtex@apr\@arrayparboxrestore
\def\@arrayparboxrestore{\revtex@apr\raggedright\let\\\tabularnewline}
\makeatother
\begin{ruledtabular}
\footnotesize
\begin{tabular}{p{0.40\columnwidth}p{0.50\columnwidth}}
Poling period $\Lambda_{\rm QPM}$  & at the group-matched retracing point (type-0, $25\,^{\circ}$C) \\
Crystal length $L$                 & $2.55$\,mm \\
Pump                               & $660$\,nm \\
Group-matched center               & $797$\,nm $\leftrightarrow$ $3.85\ \mu$m \\
Signal window (phase-matching FWHM) & $782$--$816$\,nm ($34$\,nm) \\
Idler window                       & $3.45$--$4.23\ \mu$m (${\sim}0.8\ \mu$m) \\
Reference walk-off $\delta$        & $-0.302$\,ps/mm $\to$ $\delta L = -0.77$\,ps total \\
Oven spec (drift of $\varphi_{\rm bir}$) & $-40$\,mrad/K $\to$ $25$\,mK/mrad \\
Spectrometer resolution            & required $\lesssim 25$--$30$\,pm for the ladder below (Sec.~\ref{sec:chirp}); ours $20$\,pm $\to$ $N_{\rm slot} \approx 1700$ slots/trace \\
Delay bin (peak width)             & $\approx 63$\,fs ($19\ \mu$m path) \\
Nyquist ceiling                    & $\approx 53$\,ps ($16$\,mm) \\
Delay plate (single pass)          & $T = 5.77$\,ps ($9.5$\,mm calcite) $\to$ $d_1 = 5$\,ps \\
Carrier placement (example)        & F5 at $5$\,ps, F1 at $\dT = 15$\,ps, F2 and F3 at $20$\,ps, F4 at $25$\,ps \\
Carrier group-delay spread across the window (crystal dispersion) & $0.36$ / $1.9$ / $3.4$\,ps on F1 / F2,F3 / F4; calibrated, not compensated (Sec.~\ref{sec:chirp}) \\
Spectral points per element        & ${\sim}80$ ($0.43$\,nm signal / $10$\,nm idler) \\
Tolerated sample DGD [Eq.~(\ref{eq:dgd})] & $\tau_{\rm DGD} \le 2$\,ps at this placement; $\le 4$\,ps with $\dT \approx d_1$ (Sec.~\ref{sec:tolerance}) \\
\end{tabular}
\end{ruledtabular}
\end{table}

Table~\ref{tab:numbers} translates the scheme into the parameters of the source in
our laboratory, with the QPM numerics collected in App.~\ref{app:sellmeier}. The poling
period is chosen to place the source at the signal--idler group-velocity-matched
(hereafter group-matched)
point. The mismatch curve retraces, with zero crossings at $787$ and $808$\,nm and
$|\Delta k| < 1.1\times10^{-3}\ \mu$m$^{-1}$ across the window. Phase matching
is therefore intrinsically broadband: this is the group-matched QPM design
introduced in Ref.~\cite{vanselow2019}, which underlies the broadband
mid-IR OCT source of Ref.~\cite{vanselow2020}. The phase-matching FWHM of
the bare $2.55$\,mm crystal spans $782$--$816$\,nm, and we take this as the readout window.
Group matching also makes the internal
signal--idler walk-off in the generation polarization negligible at band center
($+13$\,fs/mm, about half a delay bin over the crystal). Away from band center the
walk-off is not constant, because the idler runs from $3.5$ to $4.2\ \mu$m toward the
infrared edge of KTP; that chirp is the subject of Sec.~\ref{sec:chirp}.
The spectrometer supplies $N_{\rm slot} \approx 1700$ slots per trace. The delay window spans
three orders of magnitude between peak width (63\,fs) and Nyquist ceiling (53\,ps),
which comfortably hosts the four distinct carriers plus guard bands. The engineered splitting $d_1$
comes from a double-passed birefringent delay plate in the signal arm,
$9.5$\,mm of calcite for $d_1 = 5$\,ps (App.~\ref{app:transits}). Demodulation with a
$\pm2.5$\,ps envelope window leaves ${\sim}80$ spectral points per Jones element.
Translated to the idler via the Fourier lever
$(\lambda_i/\lambda_s)^2 \approx 23$ at the group-matched center, this is
\emph{mid-IR polarimetry across $3.5$--$4.2\ \mu$m, sampling $M(\lambda)$ at
${\sim}10$\,nm per Jones element, detected entirely on a silicon camera}. These idler
numbers ($3.5$--$4.2\ \mu$m, $10$\,nm per element) are the signal-side time--bandwidth
budget relabeled by energy conservation: they say how finely the sample is probed, not what the spectrometer resolves.

Near the resolution limit the instrument visibility rolls off, and more pixels mean
fewer photons per pixel against read noise. The read-noise side enters the Fisher
budget below (App.~\ref{app:fim}). The roll-off is quantifiable and harmless. A
$20$\,pm Gaussian instrument function multiplies whatever fringe visibility the source
delivers by $0.93$, $0.88$, and $0.82$ on the carriers at $15$, $20$, and $25$\,ps (the
finer the fringe, the more a pixel averages over it), degrading the per-parameter
Cram\'er--Rao bounds by $10$--$25\%$ (computed from the full marginalized
information matrix). Because the loss is deterministic and family specific, the
calibration frame absorbs it into the instrument contrasts: the readout stays
unbiased and continues to attain the (degraded) bounds in the Monte Carlo of
Sec.~\ref{sec:noise}.

\subsection{Crystal dispersion and calibrated demodulation}
\label{sec:chirp}

The carrier ladder of Table~\ref{tab:families} is built from group delays, and group
delays are band-center numbers. Across the readout window the idler runs from $3.5$ to
$4.2\ \mu$m, where KTP approaches its infrared absorption edge, so each crystal transit
adds to a family not only a delay but a chirp, $\varphi^{\rm nl}_\mu(\Omega)$
[Eq.~(\ref{eq:chirpladder})]. The families inherit the chirps on the same arithmetic
ladder as their delays: F1 carries $\varphi^{\rm nl}_z$, F2 and F3 carry
$\varphi^{\rm nl}_z + \varphi^{\rm nl}_y + \varphi^{\rm nl}_T$, and F4 carries
$\varphi^{\rm nl}_z + 2\varphi^{\rm nl}_y + 2\varphi^{\rm nl}_T$. Here
$\varphi^{\rm nl}_T$ is the differential chirp of the delay plate; the $HH$ element
runs along one plate axis on both passes and carries none.
Dispersion shared by all probe amplitudes, such as the ordinary-axis dispersion of the
plate, rides on all four families equally. Almost all of the chirp ($94\%$) comes
from the idler, which sits deep in the infrared dispersion of KTP; the signal at
$800$\,nm contributes only $4\%$ and the delay plate the rest. Across the window it
spreads the group delay of a
family by about $0.4$\,ps (F1) to $3.4$\,ps (F4) (App.~\ref{app:transits}).

Two consequences follow. Cosmetically, a plain Fourier transform shows $1.5$--$3$\,ps
humps instead of transform-limited peaks on F2--F4 [Fig.~\ref{fig:demo}(b)].
Substantively, the chirp must be known to the demodulation: with the four chirp
functions supplied as calibration, the matched demodulation of Sec.~\ref{sec:recipe}
collapses each family back onto the transform limit, and the bounds, the cross-talk
and the Monte Carlo bias are identical to the chirp-free case, whereas an estimator
that ignores the chirp is biased by $5$ to $160$ standard errors in the first frame.
Calibrating is free: the chirps sit on the arithmetic ladder,
$\varphi^{\rm nl}_{\rm F2} = \varphi^{\rm nl}_{\rm F3} = \tfrac12(\varphi^{\rm nl}_{\rm F1}
+ \varphi^{\rm nl}_{\rm F4})$, so any frame on which F1 and F4 are bright
determines all four chirps without Sellmeier input. The known-channel frame of
Sec.~\ref{sec:recipe} is such a frame and fixes the chirps together with the three
calibration constants; an empty arm, $M = \openone$, gives the chirps alone. We therefore carry the chirp in the model
rather than compensate it: a compensating plate in the signal arm never sees the
idler and could correct only the signal's $4\%$, and an element in the idler arm would
have to be mid-IR transparent and polarization selective.

What the chirp does cost is spectrometer resolution and guard band. At the band edges
the local carrier spacing shrinks from $5$ to $3.9$\,ps, and the chirped families must
fit the Nyquist range with their full spread. A scan of the instrument resolution with
matched demodulation keeps the per-parameter bounds within a factor $1.5$ of the ideal
readout up to $25$\,pm, $1.8$ at $30$\,pm and $3$ at $40$\,pm; beyond $57$\,pm this
ladder is unreadable. The required spectrometer resolution is therefore
$\lesssim 25$--$30$\,pm for this ladder, with $20$\,pm comfortable. At the budgeted
$N_\lambda = 80$ the ladder cannot be compressed to rescue a coarser instrument (a
scale factor of $0.7$ already raises the cross-talk by $60\%$, and $0.5$ more than
triples it); a coarser instrument instead
trades spectral points: a $110$\,pm spectrometer would need a $3$/$4$/$5$\,ps ladder
with about $16$ spectral points per element; a sample tolerance of
$0.3$\,ps then requires the repacked ladder of Sec.~\ref{sec:tolerance}.

The demodulation reads every family phase against the frequency axis of its
spectrometer, so the registration of that axis is part of the calibration, and it is
one of the quantities the fold does not make common mode (Sec.~\ref{sec:folded}). A shift $\Omega_{\rm reg}$ of the axis
moves the phase of a family at delay $\Delta_f$ by
$\Omega_{\rm reg}\,\Delta_f$; a scale error is harmless, because it shows up as a shift of
the carrier peaks, which the ladder pins. At the $20$\,pm pixel of Table~\ref{tab:numbers}
($9.4$\,GHz at $800$\,nm), one pixel is $0.29$\,rad on a phase difference within a
port (F2$-$F1, $d_1 = 5$\,ps). If the two spectrometers drift independently, a
cross-port difference sees the full $\Omega_{\rm reg}\,\Delta_f$ of each family,
$0.9$--$1.5$\,rad per pixel for the $15$--$25$\,ps of F1 and F4. A constant
offset is absorbed by the calibration frame; what has to be held or tracked between
calibration and measurement is the drift. Matching the $0.38^\circ$ bound on
$\Delta_{\rm ell}$ (F1$-$F4) of Table~\ref{tab:polarimetry} therefore asks for the
axis to be held to $0.005$ of a pixel.
Jones-matrix OCT, which faces the same requirement, meets it by tracking a
calibration signal recorded in every A-line by cross-correlation \cite{ju2013}; here a
lamp line inside the window, masked before demodulation, with ${\sim}10^4$ photons per
frame and spectrometer,
locates the axis to that precision (the centroid of a $20$\,pm line with $10^4$
photons is known to $0.09$\,pm). Reading both ports on one spectrometer, as two traces
of one camera behind a Wollaston prism, would remove the independent drift and reduce
the cross-port sensitivity to at most $\Omega_{\rm reg}\,2d_1$, twice the intra-port value,
with F3$-$F2 becoming exactly insensitive; it would not remove the need for the
line.

The accuracy of the three Jones phases rests on two calibration constants, $\varepsilon$
and $\theta{+}\kappa$ (App.~\ref{app:rates}): in the phase basis of Sec.~\ref{sec:fim}
they enter as $(\varepsilon,\ \varepsilon - (\theta{+}\kappa),\ 2\varepsilon - (\theta{+}\kappa))$,
so an error in $\varepsilon$ propagates one to one into the first two phases and doubled
into the third, and an error in $\theta{+}\kappa$ one to one into the last two. Neither
constant requires a \emph{calibrated} object. The two off-diagonal families share their
carrier,
and their phase difference F3$-$F2 equals $\arg\Mvh - \arg\Mhv - (\theta{+}\kappa)$; for
any reciprocal object that lights F2 and F3 (Sec.~\ref{sec:noise}), the first two terms
cancel, so a frame taken on it returns $\theta{+}\kappa$. The empty arm, $M = \openone$,
is achromatic and returns $2\varepsilon - (\theta{+}\kappa)$ through F4$-$F1 (the sign
of $w_4$ in Table~\ref{tab:families} is part of the known weights), hence
$\varepsilon$. What the calibration object has to be is reciprocal and off-diagonal;
its retardance, axis, and
dispersion need not be known, so an uncharacterized wave plate suffices. Using the
sample itself is possible for a sample known to be reciprocal, at the price of the
reciprocity check of Sec.~\ref{sec:demo}, which that phase difference otherwise
provides. Both constants are
phase differences of bright families and carry the statistical uncertainty of the
frame that determines them, which the numbers of App.~\ref{app:fim} do not yet
include. Being cross-port differences, they are also exposed to the independent
registration drift of the paragraph above, a systematic that is harder to control than
the statistics.

\subsection{Sample tolerance}
\label{sec:tolerance}

The single-frame readout places one demand on the sample that a sequential scan
does not. Sample thickness as such is irrelevant: an isotropic group delay sits in the
common round-trip factor and drops out of every family. Sample dispersion, likewise,
is part of the result rather than a constraint: the isotropic spectral phase of the
sample is common to all four families and, absent from the calibration frame, survives
the demodulation as the common spectral phase of the retrieved $M(\lambda)$. It only
has to fit the demodulation window. Across the idler window and in double pass, the
group delay spreads by $0.018$\,ps per mm of ZnSe, $0.081$\,ps per mm of Si and
$0.23$\,ps per mm of Ge, so the $1$\,ps guard admits some $56$, $12$ and $4$\,mm,
respectively.
What matters is the sample's
\emph{differential} group delay $\tau_{\rm DGD}$ between its polarization eigenaxes at the
idler wavelength. In double pass it splits each family into a cluster of width
$2\tau_{\rm DGD}$ in delay space, and that cluster must fit between neighboring
carriers. With a guard of $1$\,ps for crystal chirp and window edges, the rule is
\begin{equation}
2\,\tau_{\rm DGD} \le d_1 - 1\,\mathrm{ps} .
\label{eq:dgd}
\end{equation}
The largest admissible $d_1$ follows from the highest usable carrier
$\tau_{\max}$, set by the instrument visibility ($\ge 0.8$ for a Gaussian
instrument function): with F1 one carrier spacing above the DC hump, $\dT \approx d_1$, the
ladder reaches $\tau_{\max} = 3d_1$. The tolerance is then linear in the
spectrometer resolution:
\begin{center}
\footnotesize
\begin{tabular}{lcccccc}
resolution & $5$\,pm & $10$\,pm & $20$\,pm & $30$\,pm & $57$\,pm & $110$\,pm \\
\hline
$\tau_{\max}$ & $107$\,ps & $53$\,ps & $27$\,ps & $18$\,ps & $9.4$\,ps & $4.9$\,ps \\
$\tau_{\rm DGD}$ (single pass) & $17$\,ps & $8$\,ps & $4$\,ps & $2.5$\,ps & $1.1$\,ps & $0.3$\,ps \\
\end{tabular}
\end{center}
The placement of Table~\ref{tab:numbers} sets $\dT = 3d_1$ and uses only
$\tau_{\rm DGD} \le 2$\,ps of the $4$\,ps available at $20$\,pm; repacking the ladder
to $\dT \approx d_1$ recovers the factor of two. For orientation, group
birefringence estimated from phase values gives $0.27$\,ps for $10$\,mm of sapphire,
$0.4$\,ps for $10$\,mm of MgF$_2$, $1.2$\,ps for $5$\,mm of LiNbO$_3$, $2.5$\,ps for
$3$\,mm of TiO$_2$, and $3.3$\,ps for $5$\,mm of YVO$_4$. At $20$\,pm almost
everything short of thick high-birefringence crystals qualifies; at $110$\,pm only the
sapphire class does. The levers, in order, are the spectrometer resolution
($\tau_{\rm DGD} \propto 1/\delta\lambda$), the ladder repacking (factor $2$), a longer
signal wavelength ($\tau_{\max} \propto \lambda_s^2$ at fixed $\delta\lambda$, a factor
$1.6$ at $1000$\,nm within the silicon-camera range, at the price of Fourier lever and
mid-IR reach), and, beyond the tolerance, the fall-back to a sequential
scan~\cite{rajeev2026}, which has no such limit. Sharp absorption lines (gases) are a
separate limit of the same kind: they act as long delay tails, independent of thickness.

\section{Fisher information and design}
\label{sec:fim}

The preceding sections say what a single frame contains and how it is read. This
section asks how well: how precisely the seven channel parameters can be estimated
from a given number of detected photons, and how that precision depends on the
choices made above. The Fisher information is the natural currency for the question.
It turns a noise model and the fringe rates of Eq.~(\ref{eq:master}) into a lower
bound on the uncertainty of every parameter, the Cram\'er--Rao bound
\cite{helstrom1976,paris2009}, and thereby puts
different measurement schemes on one scale. We use it in four steps. We first fix
the noise model and show that the single-frame readout actually attains the
resulting bound (Sec.~\ref{sec:noise}). We then ask what
the readout costs relative to what the quantum state could deliver
(Sec.~\ref{sec:qfi}). This is where the idler, undetected everywhere else in this
paper, enters as a reference: the information available with the idler detected is
the ceiling against which the price of non-detection is measured. We then compare the
single-frame readout with the sequential polarimetry it replaces, at equal photon
number (Sec.~\ref{sec:seqvsmux}), and finally use the same
information to choose the working point (Sec.~\ref{sec:working}).

\subsection{Noise model and information budget}
\label{sec:noise}

The measurement is a spectrometer camera readout (CCD or CMOS), not photon counting.
Per pixel it gives a Gaussian count with mean $\propto I_P$ and variance
$G\,I_P + \sigma_r^2$
(shot noise at analog camera gain $G$ plus read noise; App.~\ref{app:fim}).
The Fisher information matrix (FIM) for the seven channel parameters
$\boldsymbol{\vartheta} = (|\Mhh|,\allowbreak |\Mhv|,\allowbreak
|\Mvh|,\allowbreak |\Mvv|,\allowbreak
\arg\Mhv{-}\arg\Mhh,\allowbreak \arg\Mvh{-}\arg\Mhh,\allowbreak
\arg\Mvv{-}\arg\Mhh)$,
which we abbreviate below by the name of the numerator alone, $\arg\Mhv$ for
$\arg\Mhv{-}\arg\Mhh$ and so on,
follows from Eqs.~(\ref{eq:master})--(\ref{eq:phases}) with the global interferometer
phase $\Phz$ (and, where uncalibrated, the calibration phases $\eta_f$ of
Eq.~(\ref{eq:phases}), the fixed crystal and plate phases of each family) as nuisance
parameters.

The resulting classical bound is not merely a bound: the single-frame readout of
Sec.~\ref{sec:recipe} attains it. In a Monte Carlo of $600$ simulated frames of the
channel of Fig.~\ref{fig:demo} ($N = 10^6$ detected photons per frame, Poisson pixel
noise, $\Phz$ estimated alongside and marginalized) the standard deviation of every
parameter matches its Cram\'er--Rao bound to within
$\sigma_{\rm MC}/\sigma_{\rm CRB} = 0.97$--$1.05$, with no resolvable bias, as expected
for the least-squares readout at these counts per pixel, where the Poisson likelihood
is close to Gaussian. This survives the
instrument function: with the $20$\,pm resolution of Sec.~\ref{sec:feasibility} folded
into simulation and calibration, the readout remains unbiased and matches the
correspondingly degraded bounds. The hierarchy of the next subsection therefore starts
from an operationally attained level. One class of channels is excluded from these
statements: a sample diagonal in the $H/V$ basis, such as a wave plate aligned with the
axes, leaves F2 and F3 dark. Its two off-diagonal phases are then undefined and the
seven-parameter bound diverges, as it does for any polarimeter analyzing a sample in
the polarimeter's own eigenbasis.

\subsection{Quantum limits and the price of non-detection}
\label{sec:qfi}

Before comparing readout protocols (the spectrometer readout of
Sec.~\ref{sec:readout} against sequential schemes, Sec.~\ref{sec:seqvsmux}) we ask
how much information the state itself carries. Because the continuous-wave (CW) pump
fixes the idler detuning to $-\Omega$ for a signal photon at $\Omega$, any measurement
on the detected beam sees the pairs at different $\Omega$ as an incoherent mixture:
the signal state is block diagonal in $\Omega$, and its Fisher information is a sum
over frequency bins (the idler-inclusive state keeps the coherence between bins, a
point we return to in App.~\ref{app:fim}). Within each bin the state has
the form $\sum_{ab} \mathcal{A}_{ab}\,\ket{s_a}\ket{e_b}$, where $a \in \{H,V\}$
labels the detected signal polarization and $b$ runs over the four undetected modes:
the two idler polarizations and the two environment modes into which the channel
absorbs or scatters (App.~\ref{app:rates}). The $2\times4$ amplitude matrix
$\mathcal{A}(\Omega)$ determines everything that follows: the reduced signal state
$\mathcal{A}\mathcal{A}^\dagger$ gives the information available without the idler,
the idler-inclusive state gives the information with it. The quantum Fisher
information (QFI) \cite{braunstein1994,paris2009} follows from the
symmetric-logarithmic-derivative (SLD) equation on each block (App.~\ref{app:fim});
for the seven-parameter problem it is a matrix \cite{liu2019qfim}, and we compare
determinants \cite{kiefer1960}. This gives a three-level
hierarchy at equal photon number, all with $\Phz$ marginalized. The three levels are the
classical information of our actual readout, $F_{\rm spec}$; the QFI of the signal alone
(all measurements on the detected beam, idler and environment traced), $F_{\rm sig}$;
and the QFI with the idler detected as well, $F_{\rm sig+idl}$. For the example
channel,
\begin{equation}
\Big(\tfrac{\det F_{\rm spec}}{\det F_{\rm sig}}\Big)^{1/7} = 0.45 ,
\qquad
\Big(\tfrac{\det F_{\rm sig}}{\det F_{\rm sig+idl}}\Big)^{1/7} = 0.19 .
\label{eq:hierarchy}
\end{equation}
The determinant serves as a scalar measure of the joint information on all seven
parameters, and the seventh root turns each ratio into a geometric mean per parameter
(per-parameter uncertainties scale as the inverse square root
of the information, $\sigma \propto 1/\sqrt{F}$, so each information ratio quoted here enters
the uncertainty as its inverse square root). Two conclusions follow. First, the
spectrometer readout is not optimal (that this gap is real, and not an artifact of an
unattainable quantum bound, is shown below): a factor
$1.39$--$1.61$ across the seven parameters (Table~\ref{tab:perparam};
$0.45^{-1/2} = 1.5$ on average) separates
it from the best measurement on the detected beam. This gap is not a property of the
fixed $H/V$ analysis basis. Evaluating the same Poisson model for other polarization
measurements per spectral bin gives a readout efficiency of $0.454$ for the best
fixed rotated basis and $0.457$ for $H/V$ analyzed against an optimized second basis,
against $0.452$ for the $H/V$
readout: the choice of basis gains at most about one percent. Information is lost
instead by a three-basis polarimeter (linear $H/V$, linear at $\pm45^\circ$, and
circular) and by a symmetric
four-outcome measurement ($0.441$ and $0.439$), so a polarimeter in front of the
spectrometer does not close the gap. The gap is the generic price of
reading amplitude and phase from a scanned fringe rather than from the optimal
measurement. For a single fringe channel of visibility $0.7$--$0.9$, the scanned
readout retains $0.30$--$0.42$ of the quantum information on the modulus
[Eqs.~(34) and (48) of Ref.~\cite{panda2026}] and, evaluating the same fringe integral
for the phase against Eq.~(35) there,
$0.58$--$0.70$ on the phase, a
geometric mean of $0.46$--$0.49$, just above our $0.45$.
Second, the price of not detecting the mid-IR photon is a factor between $2.3$ and
$2.4$ in uncertainty in the determinant sense, that is, on $(\det F)^{1/14}$:
$0.19^{-1/2} = 2.3$ when the frequency bins of
the idler-inclusive state are treated independently, which bounds $F_{\rm sig+idl}$
from below, and $2.4$ against the fully coherent purified state, which bounds it
from above (App.~\ref{app:fim}). The individual parameters spread from $1.8$ to $2.7$
around a geometric mean of $2.16$, somewhat larger on the
moduli (median $2.2$) than on the phases (median $2.0$), Table~\ref{tab:perparam}:
with the idler detected, the moduli become accessible through the idler's
polarization tags, the phases only through interference. That is a quantitative version of the
central promise of measurement with undetected photons. For single-parameter
absorption estimation the corresponding loss of quantum Fisher information upon
tracing out the idler is known in closed form for the induced-coherence
configuration [Eq.~(56) of Ref.~\cite{houde2026}]; what is new here is the
two-sided bracketing of the multiparameter price for a polarization channel.

Multiparameter quantum Cram\'er--Rao bounds need not be jointly attainable
\cite{ragy2016,albarelli2019}, so we quantify the
incompatibility by the measure $R = \lVert i F^{-1} \mathcal{D} \rVert_\infty$
\cite{carollo2019,albarelli2019},
the largest eigenvalue modulus of $i F^{-1}\mathcal{D}$, where $F$ is the SLD quantum Fisher
matrix of the model at hand, the SLD operators $\hat L_a$ carry the hat to
distinguish them from the crystal length $L$, and
$\mathcal{D}_{ab} = \operatorname{Im}\operatorname{Tr}[\rho \hat L_a \hat L_b]$ [Eqs.~(26)--(27) of
Ref.~\cite{albarelli2019}; twice the mean Uhlmann curvature of
Ref.~\cite{carollo2019}] (the symbol $R$ here is unrelated to the reflection $R(A)$
of Sec.~\ref{sec:rates}). Here $R = 0$ means compatible, and $R \le 1$
always; the Holevo bound \cite{holevo1982} exceeds
the SLD bound by at most a factor $(1 + R)$ [Eq.~(25) of Ref.~\cite{albarelli2019}].
At the example channel, with $\Phz$ marginalized,
$F_{\rm sig}$ is nearly compatible ($R = 0.07$). The largest normalized pairs
$|\mathcal{D}_{ab}|/\sqrt{F_{aa}F_{bb}}$, between amplitudes and between amplitude and phase of
individual elements, all stay below $0.04$. The Holevo bound, attainable
asymptotically with collective measurements \cite{albarelli2019}, thus exceeds the
signal-alone SLD bound by at most $7\%$ for any weight matrix, so the factor-$1.5$
readout gap above is a property of the readout, not of the bound.

$F_{\rm sig+idl}$, by contrast, is strongly incompatible ($R = 0.74$), as expected
for a pure-state limit; the incompatibility is dominated by the amplitude--phase
pairs of individual matrix elements, the same pairing that makes phase and loss
incompatible in a detected interferometer \cite{crowley2014}. Since $\mathcal{D} \neq 0$, the SLD bound at that level is not
attainable for every weight matrix, even asymptotically \cite{ragy2016} [Eq.~(36) of
Ref.~\cite{albarelli2019}], so part of the nominal $2.3$--$2.4$ price of
non-detection is an artifact of the bound rather than physics. How much is not fixed
by $R$, which only caps the gap, but by the Holevo bound itself.
Evaluated as a semidefinite program \cite{albarellifriel2019} [Eqs.~(19)--(22) of
Ref.~\cite{albarelli2019}] with the signal-alone QFI as weight matrix, it lowers the
scalar price of non-detection from $2.15$ (SLD) to $1.96$: a $9\%$ gap despite
$R = 0.74$.

These statements are generic rather than properties of the example channel. Over
$20$ random contractions the readout efficiency stays within $0.44$--$0.50$ and
the price of non-detection within $0.18$--$0.27$. Both ratios track the total
channel transmission $\operatorname{Tr}(M^\dagger M)$ (rank correlation $-0.89$ and
$-1.00$ over the ensemble, lossier channels at the upper end of each range) and are
nearly insensitive to how unequally the channel transmits the two polarizations
(the ratio of its singular values; rank correlation $-0.19$ and $-0.32$).
The direction is the one known from loss estimation, where the information a detected
probe carries about a transmission grows without bound as the channel becomes
transparent \cite{monras2007,adesso2009}: the more transmissive the sample, the more
the undetected idler costs.
$R_{\rm sig}$ never exceeds $0.15$, so the signal-alone bound remains essentially attainable across
the ensemble. By contrast, $R_{\rm sig+idl}$ stays above $0.8$ for every random
channel. The example channel, at $0.74$, is the \emph{most favorable} case in the
whole set, so for generic, lossier channels the conclusion that part of the
non-detection price is a bound artifact only sharpens.
The $\chi = 0$ state limit passes the same construction as a cross-check. A purified
state model (idler $\otimes$ dephasing environment $\otimes$ loss modes) reproduces
the $\chi \to 0$ limit of the process QFI to numerical precision, at the signal
level exactly for \emph{all} values of $\purity$. The reason is that the signal's reduced state
feels $\purity$ only through the fringe amplitude (App.~\ref{app:fim}). The state-case
hierarchy at the mixed benchmark state reads
$(\det F_{\rm spec}/\det F_{\rm sig})^{1/3} = 0.57$ and
$(\det F_{\rm sig}/\det F_{\rm sig+idl})^{1/3} = 0.21$.

\subsection{Sequential against delay-multiplexed at equal budget}
\label{sec:seqvsmux}

\begin{figure}[t!]
\includegraphics[width=\columnwidth]{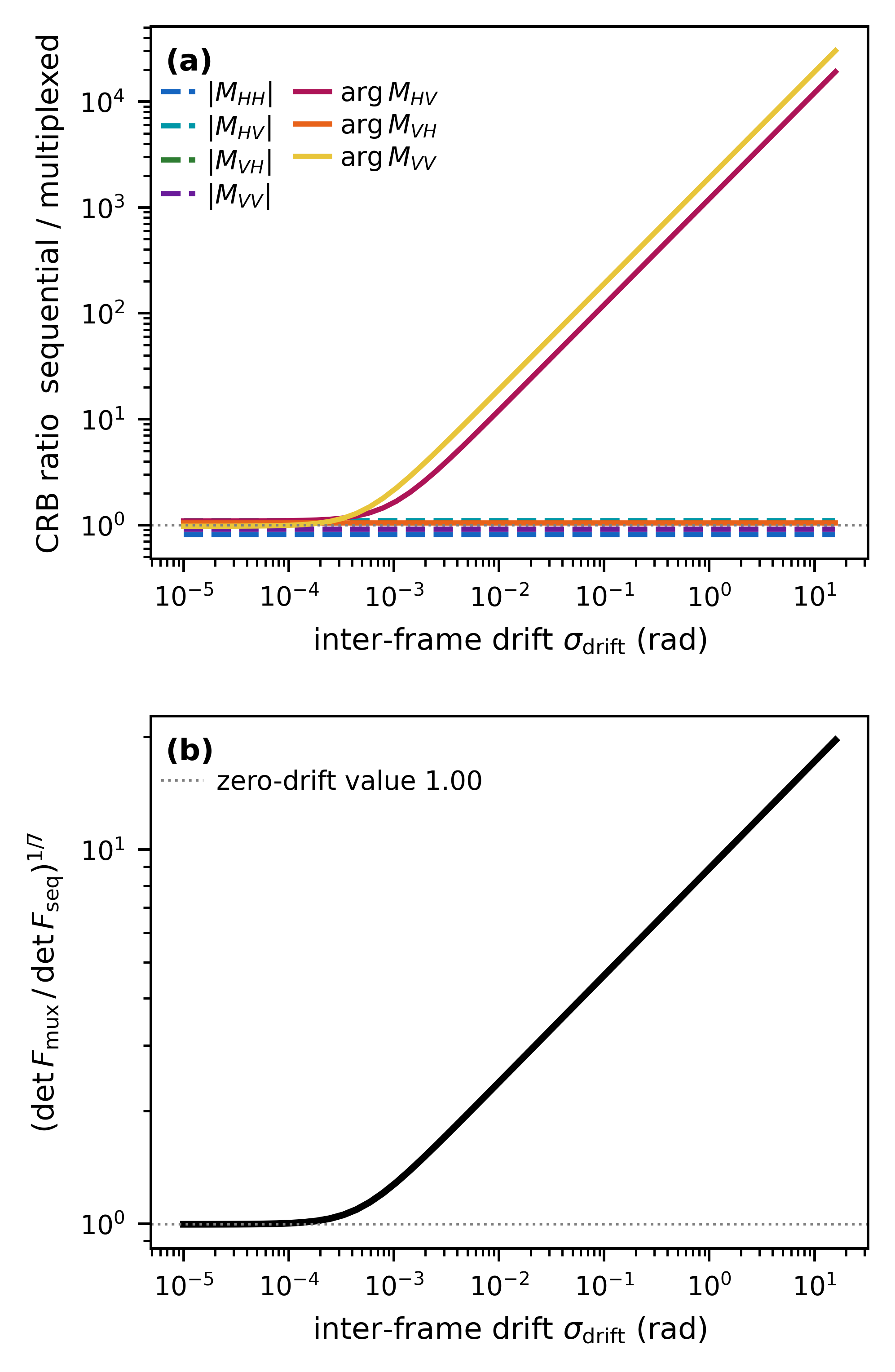}
\caption{\label{fig:detf}%
Delay-multiplexed against sequential process tomography at equal photon
budget: $N = 10^8$ detected photons per campaign, i.e.,\ $100$ frames of $10^6$, with
Poisson pixel noise and the example channel of Fig.~\ref{fig:demo}. (a)~Ratio of the
CRBs, sequential over multiplexed, for each of the seven parameters
as a function of the inter-frame phase drift $\sigma_{\rm drift}$. Beyond the
shot-noise floor (about $10^{-3}$\,rad at this budget) the sequential scheme loses
the two phases that connect the columns of $M$, their bounds growing in proportion
to $\sigma_{\rm drift}$, while the parameters read within a single frame stay near
parity. (b)~The same comparison aggregated into the geometric mean
$(\det F_{\rm mux}/\det F_{\rm seq})^{1/7}$ over the seven parameters.}
\end{figure}

A sequential protocol switches settings in the manner
of Refs.~\cite{fuenzalida2022b,rajeev2026}, at the same total photon number. The
comparison with it is fair by construction, because the total flux does not depend
on the settings [Eqs.~(\ref{eq:bgH})--(\ref{eq:bgV})]. We model the sequential
baseline as two state-mode frames ($\chi = 0$, $\chi = 90^{\circ}$) at half the
exposure each, with an inter-frame drift
$\epsilon_{\rm d} \sim \mathcal N(0, \sigma_{\rm drift}^2)$ marginalized alongside
$\Phz$ (App.~\ref{app:fim}). Figure~\ref{fig:detf} shows the result. At
zero drift the two schemes are on par,
$(\det F_{\rm mux}/\det F_{\rm seq})^{1/7} = 1.00$ for the example channel, so
multiplexing four families into one frame is essentially information-neutral. Per
parameter the zero-drift comparison is a mild trade. The sequential frames measure
one column of $M$ at a time (F1, F3 at $\chi = 0$; F2, F4 at $\chi = 90^\circ$) and
so concentrate their weight on fewer families (better on the diagonal moduli, ratios
$0.81$ and $0.91$). The multiplexed frame wins on the off-diagonal elements
and their phases ($1.05$--$1.10$), with $\arg \Mvv$ near parity ($0.97$). Beyond the
shot-noise floor the two schemes diverge. The sequential scheme reads the
inter-column phase as a difference
between two \emph{time-separated} frames, so the uncertainty of the two phases that
connect the columns grows linearly with $\sigma_{\rm drift}$: already at
$\sigma_{\rm drift} = 1$\,rad the sequential bounds on $\arg\Mhv$ and $\arg\Mvv$ are
$1.2\times10^3$ and $1.9\times10^3$ times the multiplexed ones, while the other five
parameters stay within the zero-drift ratios above. The multiplexed frame, by contrast,
measures all relative phases instantaneously. Drift within the exposure is a different
matter and affects both schemes alike: since all four bright families carry $\Phz$
with the same sign, it damps their moduli by one common factor and leaves the phases
untouched. Unlike the instrument function, which is deterministic and family specific,
it fluctuates from frame to frame and therefore appears as extra scatter of the moduli
rather than as a calibrated contrast. The flat
direction of the sequential Fisher matrix is exactly the drift mode. The
single-shot capability thus costs no information and buys drift immunity. This
drift advantage is a property of the baseline as much as of our scheme: it holds
for sequential protocols that carry no phase reference inside each frame and
connect their frames through an external phase standard, which is the situation of
the protocols cited above. A sequential protocol that co-measures a reference
family in every frame would remove the drift term at the cost of photon budget.
Giving up the phase reference altogether, and working from visibilities alone, costs
a third measurement round instead \cite{kysela2024}.

Both conclusions are generic, not artifacts of the example channel. Over $40$
random contractions the zero-drift ratio clusters at parity (median $1.00$, range
$0.98$--$1.06$), while already $1$\,mrad of inter-frame drift shifts the whole
distribution in favor of multiplexing (median $1.12$, up to $1.38$). Camera read
noise adds an independent argument, the Fellgett situation for polarimetric
multiplexing: no advantage under Poisson noise, an advantage once detector noise is
present. At fixed pixel pitch, the sequential protocol needs two frames and reads the camera
once per frame, so it pays the read-noise dose twice. At the single-frame budget ($10^6$ photons, $15\,e^-$ read
noise per pixel, the range of our silicon spectrometer camera) this alone tilts the
comparison to $(\det F_{\rm mux}/\det F_{\rm seq})^{1/7} = 1.50$ at zero drift.

Two remarks put these numbers in context. First, our sequential baseline is the
\emph{strongest} sequential protocol: in each of its frames both PBS ports carry a
fringe, so no photons are burned. The protocol actually used in
Ref.~\cite{fuenzalida2022b} switches a half-wave plate in the signal arm, ahead of
the beam splitter, such that only one fringe channel is live per setting, parking
one third of the photon budget on a non-fringing detector. Against that baseline the $\chi = 0$ state frame of our
scheme gains exactly $3/2$ in the Fisher information of each visibility parameter
($\Pi_H$ and $\purity$) and a state-dependent factor $2.0$--$2.9$ in $\det F$ at zero
drift (App.~\ref{app:fim}). The multiplexed advantage would stack on top.
Second,
with pure Poisson noise all ratios are independent of the photon budget $N$ (both
Fisher matrices scale linearly in $N$). Only the read-noise argument and the
location of the drift knee (the shot-noise floor
$\sigma_{\rm shot} \propto 1/\sqrt N$) depend on $N$.

\subsection{Working-point design}
\label{sec:working}

The design knobs are $\chi$, $A$, $B$, $\dT$, and the carrier splitting $d_1$.
The delays $\dT$ and $d_1$ do not enter the Fisher matrix, to an excellent approximation, once the
carriers are resolved and clear of the Nyquist limit. They only place the carriers
(Sec.~\ref{sec:budget}), which leaves $(\chi, A, B)$ as the design space that
matters for precision.
Since the budget splits over families as $w_f^2$, optimizing
$\det F$ (or a weighted single-parameter criterion) over the knob space is a classical
optimal-design problem \cite{kiefer1960}. In detected-light polarimetry the same
philosophy underlies the variance-optimal Stokes and Mueller designs of
Refs.~\cite{sabatke2000,goudail2017,foreman2019}, and the determinant criterion
itself is established there: Ref.~\cite{foreman2008} builds the Fisher matrix for
Stokes and Mueller estimation under Poisson noise and optimizes its determinant,
which that literature calls D-optimality. What is new here is the object, a Jones matrix carried by a beam that
is never detected, not the criterion.
The result is simple enough to state without a figure. A full $(\chi, A, B)$
scan puts the global $\det F$ optimum at the balanced point
$\chi = A = B = 45^\circ$, with a broad top (a few percent over $\pm5^\circ$,
${\sim}20\%$ at $\pm10^\circ$).
Returning the pump at the polarization angle $B = A$, equal to the reflection
angle of the signal plate, is the optimum for the example channel, not just
convenient. It is optimal to within a fraction of a percent for generic channels:
the optimal return-pump angle does depend on the channel, but only weakly. Over $20$
random channels it stays
within $2.4^\circ$ of $45^\circ$, and holding $B = 45^\circ$ fixed instead of tuning
it to each channel raises every per-parameter Cram\'er--Rao bound by less than
$0.3\%$. The pump polarization can therefore stay fixed, whatever the sample.
This contrasts with the state limit, where the two ports are not
equally informative: the $H$ port carries only $\Pi_H$, while the $V$ port carries
two of the three state parameters ($\purity$ and $\xi$). The optimum therefore tilts
away from $45^\circ$ toward the $V$ port, $A^\ast \approx 46^\circ$--$51^\circ$ over
the benchmark states, and likewise $\alpha^\ast_{\rm an} \approx 49^\circ$--$51^\circ$
for the analyzer angle $\alpha_{\rm an}$ of a parallel readout in the two-source
geometry of Ref.~\cite{fuenzalida2022b}. The tilt is a parameter-count effect, not
a transmission asymmetry.
The process case probes both columns symmetrically, and
for a generic channel neither deserves extra budget. The working point is
therefore forgiving: even at a $30{:}1$ diattenuation ratio the optimum stays
within $2^\circ$ of the balanced point, and adaptive tilting buys less than
$0.5\%$ per parameter.

The balanced optimum carries a hardware dividend: the pump arm needs no
wave plate, since a bare end mirror returns the diagonal pump at $B = 45^\circ$ by
itself, up to a fixed phase that joins the calibration constants of
App.~\ref{app:rates}. The arm itself can then be dropped: the pump travels with the
signal and is returned before the delay plate by a long-pass filter,
and the fold collapses to a signal arm and an idler arm.
The price is the $\chi = 0$ state limit, which needs the pump
rotated to $B = A$ (Sec.~\ref{sec:scheme}) and hence a plate in a pump arm of its
own; the two-arm layout is tied to the process working point.

\section{Discussion and outlook}
\label{sec:discussion}

This section places the scheme in context: against its sequential counterpart,
against the boundary that limits what any low-gain measurement with undetected
photons can reconstruct, against classical mid-infrared polarimetry and the
SU(1,1) estimation literature, and along the routes that extend it, from
depth-resolved readout to magneto-optical and chiral metrology.

\paragraph{Relation to sequential process tomography.}
The sequential counterpart of this work disentangles the coherent sum over channel
elements by \emph{switching} known unitaries on the signal
photon~\cite{rajeev2026}: for a channel on a mode space of dimension $d$
($d$ orbital-angular-momentum modes on each of signal and idler in that work; $d$ is
unrelated to the rung $d_1$), the protocol applies $d(d{-}1)/2$ two-mode rotations,
each at two angles, one after the other. In their example all phases are referenced
to one interference pattern, so phase stability across the campaign is assumed
without being stated. Its reach is any dimension $d$ and arbitrary channels, including
non-unitary ones, at $O(d^2)$ interferograms. We disentangle by \emph{delay tags} instead: one frame, no
settings, all relative phases instantaneous --- but qubit polarization and a coherent
channel (7 parameters). The two designs are complementary: breadth bought with time
versus speed bought with spectrum. The fold enters the same balance:
Ref.~\cite{rajeev2026} reads a channel traversed once between two sources, whereas we
read the round-trip matrix of one arm, which for a reciprocal sample carries five
independent parameters and does not by itself give the single-pass matrix. For the
unitary case that Ref.~\cite{rajeev2026b} addresses, a benchmark at the quantum
Cram\'er--Rao bound exists for high-dimensional unitaries close to the identity
\cite{escandonmonardes2024}; the estimation theory of the present work is the
counterpart for a lossy polarization channel.

\paragraph{The moment boundary of this line of work.}
The unitarity argument of Sec.~\ref{sec:witness} generalizes to a statement about
what measurements with undetected photons can access at all. A general channel
on the idler polarization is a unitary $U$ acting on the idler and an
environment $E$ (loss modes, scattering modes) that starts in its vacuum. Every
interference term in a low-gain scheme is the overlap of two idler states, and the
two can only interfere if the environment is left in the vacuum after the passage,
otherwise its state would record which idler went through the channel. The
overlap is therefore the vacuum-to-vacuum amplitude
$K_{ji} = \langle j, \mathrm{vac}_E|U|i, \mathrm{vac}_E\rangle$,
a single $2\times 2$ matrix, linear in $U$: the channel's \emph{first
moment}, and our coherent $M$. Probe--probe pairings (F5/F6) add nothing beyond
this, since two idlers that both passed the channel overlap as
$\langle j, \mathrm{vac}_E|U^\dagger U|i, \mathrm{vac}_E\rangle = \langle j|i\rangle$,
the input overlap.

Counting parameters shows what is missing. A general completely positive,
trace-preserving (CPTP) qubit map has $12$ real parameters; $M$ carries $7$ ($8$
minus the unobservable global phase). The remaining $5$ describe decoherence: how
much of the photon went into the environment and where. (The channel here is a qubit
with leakage into $E$ rather than a trace-preserving map on the qubit alone, so the
count is indicative; what it counts correctly is the kind of information that is
missing.) They would enter only
through quantities quadratic in $U$, the ones a detected-photon transmission
measurement reports, and no low-gain interference term contains them. Loss and
dephasing therefore both shrink $|K|$ and are indistinguishable within a
frame. This limit is not specific to our readout. Switching known signal
unitaries, as in Ref.~\cite{rajeev2026}, re-probes the same $K$ along
different linear combinations of its entries: the arbitrary, not necessarily
unitary, operations reconstructed there are arbitrary $M$, that is, single-Kraus channels,
and what any protocol in this class reconstructs is that coherent part.

\paragraph{Resolution moves that boundary.}
The boundary is relative to a \emph{resolution cell}, however, and this is where a
single-shot, spectrally resolved protocol changes the accounting. Much of what
registers as decoherence is not intrinsic but an averaging artifact: the channel
depolarizes because $M$ varies over some coordinate (wavelength, time, position,
transverse mode) that the measurement integrates over. What a frame returns is the
cell average $\langle M \rangle$, and the missing parameters are exactly the variance
$\langle M \otimes M^* \rangle - \langle M \rangle \otimes \langle M \rangle^*$ across
that cell. Shrinking the cell therefore converts decoherence into resolved variation.
Delay multiplexing shrinks it along two axes at once. \emph{Spectrally}: demodulation
returns $M(\lambda)$ at ${\sim}80$ points rather than one matrix per campaign
(Sec.~\ref{sec:budget}). Spectral dephasing then appears as a wavelength dependence of
$\arg M$ instead of as an unexplained loss of modulus. \emph{Temporally}: the
resolution time is the exposure and not the campaign, so $K$ becomes a time
series. Channel fluctuations slower than the frame rate then appear as scatter of the
per-frame estimates rather than as mean shrinkage. Once the averaging contribution is
resolved away, the residual shrinkage of $|K|$ is unambiguously loss --- the
degeneracy that closes the single-frame accounting is opened by resolution rather than
by extra settings. What stays genuinely invisible is decoherence below \emph{all} of
the protocol's resolutions: faster than one frame, finer than one spectral bin, or
across the transverse mode.

Two routes reach second moments proper. (i)~\emph{Double-pass moments}: the folded
geometry measures $\langle M_{\rm back} M_{\rm forth}\rangle$, a second moment for
disorder that is static within a shot. Compared with an independently calibrated
single-pass $\langle M\rangle$, it separates loss from static dephasing: a
single-pass visibility $\mathcal{V}$ returns as $\mathcal{V}^2$ for loss but as
$\mathcal{V}^4$ for a Gaussian
random phase $\zeta$, since the phase adds on the two passes
($\langle e^{2i\zeta}\rangle = \langle e^{i\zeta}\rangle^4$). (ii)~\emph{High parametric gain}: the returning idler
is then amplified rather than merely re-emitted, the channel enters the signal
intensity twice, and count correlations carry $M^\dagger M$ directly. A four-mode
Gaussian estimate for a two-pass geometry puts the onset at moderate gain. With the
first-moment visibilities matched, count covariances separate loss from
shot-to-shot dephasing at $3\sigma$ within ${\sim}10^{3}$ frames already at
$\sinh^{2}\!g \approx 0.1$--$0.2$, and within ${\sim}10^{2}$ frames near
$\sinh^{2}\!g \approx 1$. Here $g$ is the parametric gain in the sense of the squeezing
parameter ($\sinh^{2}\!g$ = mean photon number per mode), not the camera gain $G$
of Sec.~\ref{sec:noise}. In the low-gain
limit, by contrast, the distinguishing signature lives in the count covariance and
falls as $g^{4}$, while the first-moment information falls only as $g^{2}$ with the
photon number; per detected photon it vanishes as $g^{2}$.
A polarization-resolved nonlinear interferometer has already been extended
theoretically into the high-gain regime in Ref.~\cite{oglialoro2025}, where interference of the idler at the
wave plate adds terms with no low-gain counterpart; whether the carrier ladder
survives them is open.

\paragraph{Depth-resolved Jones matrices: PS-OCT with undetected photons.}
If the idler arm contains a layered, reflective sample, the returning idler is a sum
over layers $\sum_l r_l M_l$ with round-trip delays $\tau_l$. Every family peak
becomes an A-scan comb (replicas at $\Delta_f + \tau_l$), and each layer carries its
own Jones matrix: PS-OCT of an undetected beam, in a single frame.
Axial resolution is $\lambda_i^2/(2\Delta\lambda_i) \approx 10\ \mu$m at the full
$0.8\ \mu$m idler bandwidth. The depth range is the carrier spacing minus the guard
band: at the placement of Table~\ref{tab:numbers} about $4$\,ps round trip, or
$0.6$\,mm of optical path, which accommodates the few-interface samples of mid-IR
OCT (paint layers, ceramics \cite{vanselow2020}) with room to spare; a finer spectrometer
widens it linearly (Sec.~\ref{sec:tolerance}). Depth and spectral
resolution of $M$ are Fourier conjugates of the same axis, so the depth axis comes
for free rather than as an extra one, but depth bins and spectral points share the
demodulation window: a four-layer sample retains about $20$ spectral points per
layer. The matched demodulation of Sec.~\ref{sec:chirp} survives the comb: in a
simulation with four non-dispersive interfaces spread over $2.4$\,ps, the replicas of
every family share its chirp and sit fifteen peak widths apart, so one calibration
set collapses the whole comb. The per-parameter bounds stay within $5\%$ of the
single-interface values on all families and layers (worst case $1.05$), cross-talk
remains below $0.3\%$, and the Monte Carlo shows no bias resolvable at $3\sigma$ in
$300$ frames. The price is the visibility roll-off across the F4 comb
($0.82 \to 0.79$).
Layer--layer autocorrelation terms
(interference of one first-pass component with itself across layers, which unitarity
does not forbid) require $d_1$ larger than the sample's total optical
thickness. Otherwise they contaminate the dark-carrier region and spoil the null test.
Prior art has the depth axis without polarization \cite{vanselow2020,zorin2026};
polarization-sensitive quantum OCT \cite{booth2004,booth2011,sukharenko2021,sukharenko2024} and
Hong--Ou--Mandel birefringence imaging with a Fisher analysis of the axis angle
\cite{goncalves2026} both detect the two photons in coincidence. The polarization
structure of an undetected vector beam has been reconstructed point by point
\cite{vasikonis2026}, recovering $M\ket{H}$ for a known vortex wave plate in the
idler arm from two off-axis holograms, one per polarization projection: a single
column of a Jones matrix for a calibrated element rather than a sample, with the $H$--$V$
relative phase read across the two frames, and spatially but not spectrally resolved.
Depth-resolved Jones-matrix polarimetry of an undetected beam has not been realized:
the prospect is named as polarization-sensitive OCT in
Ref.~\cite{fuenzalida2022b}, and Ref.~\cite{oglialoro2025} describes the area as
largely unaddressed.

\paragraph{Magneto-optics and a chiral mode.}
Magneto-optical metrology is a natural target for the spectropolarimeter.
A Faraday rotation $\theta_{\rm F}$
per pass is a circular retarder. On the round trip it becomes the rotation
$M = \mathcal{R}(2\theta_{\rm F})$ [with $\mathcal{R}(\cdot)$ the proper rotation
matrix, $\det \mathcal{R} = +1$, as opposed to the reflection $R(A)$ of
Sec.~\ref{sec:rates}], so $\Mhv = -\Mvh = -\sin 2\theta_{\rm F}$. It
lights up the off-diagonal families at \emph{first} order in $\theta_{\rm F}$, with an
antisymmetric phase signature that separates it from linear birefringence of the
same sample. A visibility-only readout carries $\theta_{\rm F}$ only in even powers, through
$\cos 2\theta_{\rm F}$ and $\cos\theta_{\rm F}$ respectively [Eqs.~(5) and (7) of
Ref.~\cite{chakraborty2025}], so the
change in fringe visibility scales as $\sin^2\theta_{\rm F}$ and is therefore weak at
small angles.
That is precisely the regime of thin films, small Verdet
constants, and weak fields. The folded geometry adds a selectivity dividend.
The idler double-passes the sample, so the non-reciprocal Faraday rotation
\emph{doubles} while reciprocal optical activity (cell windows, substrate)
cancels on the round trip. Reversing the magnetic field flips the sign of
$\theta_{\rm F}$ but leaves every reciprocal contribution unchanged, so the
difference of two runs is a built-in differential measurement. The wavelength
dependence of the rotation (the Verdet dispersion) arrives in the moduli of
the off-diagonal families (the round-trip rotation matrix is real) and magnetic
circular dichroism in their quadrature phase (circular diattenuation is purely
imaginary in the linear basis). Rotation and dichroism are Kramers--Kronig
partners, here measured simultaneously.
The counterpart of this selectivity is a structural blind spot. The measured $M$ is
the round trip of the idler arm, and for any \emph{reciprocal} sample the
round-trip Jones matrix is symmetric. Natural optical activity and natural
circular dichroism therefore cancel exactly (the end mirror flips the helicity).
The antisymmetric part of $M$ is thus a background-free monitor of non-reciprocity. It is
the physics behind the consistency check of Sec.~\ref{sec:demo}, where the estimator
returns $\Mhv$ and $\Mvh$ separately and a reciprocal channel makes the two agree.
A linear two-crystal arrangement, with the idler traversing the sample once between
the two crystals, would instead deliver the single-pass matrix with all seven
parameters independent, but it would lose the reciprocity filter. Reciprocal and
Faraday rotation are then one and the same Jones matrix within a frame, separable
only by a second measurement with the direction of light reversed.
This complementarity is toggled by the flip-in quarter-wave plate at $45^\circ$ in
front of the idler end mirror (Fig.~\ref{fig:setup}), whose double pass is the
exact $H \leftrightarrow V$ exchange $\sigma_x$ (App.~\ref{app:transits}); we call
it the swap plate below.
With the exchange in place a reciprocal rotation $\alpha_{\rm OA}$
survives as $M = \sigma_x \mathcal{R}(2\alpha_{\rm OA})$, while the Faraday rotation
cancels, $\mathcal{R}(\theta_{\rm F})\,\sigma_x\,\mathcal{R}(\theta_{\rm F}) = \sigma_x$. Natural circular
dichroism reappears in quadrature on the diagonal of $M$, and magnetic circular
dichroism drops to a polarization-independent loss. One removable mid-IR
wave plate thus switches the same single-frame readout between a magneto-optical
and a chiral spectropolarimeter, each with the other's circular response nulled
by geometry. The exchange also turns axis-aligned linear retardance into a global
phase, so the two configurations probe complementary slices of the sample.
Two qualifications temper the chiral mode. First, an angle error $\epsilon_{\rm sw}$
of the swap plate is \emph{exactly} degenerate with a reciprocal rotation
(apparent rotation $-\epsilon_{\rm sw}$, both being the same rotated exchange axis).
The plate-angle calibration \emph{is} therefore the zero point of the chirality
measurement: $1^\circ$ of plate error reads as $1^\circ$ of optical
activity. The empty-arm null test of App.~\ref{app:transits} measures
precisely this offset (F1/F4 residual $2\epsilon_{\rm sw}$, linear), and is
therefore constitutive for this mode rather than a convenience. Second, the
exchange is narrowband. Across the idler window ($\Delta\lambda/\lambda
\approx 0.20$, Table~\ref{tab:numbers}) a standard zero-order plate cut for
$3.85\ \mu$m retains a residual retardance up to $0.17$\,rad ($9.7^\circ$). That
lights F1/F4 at up to $0.17$ of full amplitude at the window edges (birefringence
dispersion neglected, so a lower bound). The residue is deterministic and open to
calibration, but a true null only near the group-matched center. Its cause is the
dispersion of the birefringence, so removing it means leaving birefringence behind.
A mid-IR Fresnel rhomb does that: its retardance comes from total internal
reflection and is achromatic by construction. ZnSe units specified
across $2$--$15\,\mu$m hold their retardance to $\pm2^\circ$ per pass, flat over the
entire $3.5$--$4.2\,\mu$m window; the observable residue, set by the \emph{single}-pass
retardance error $\epsilon_\Gamma$ (App.~\ref{app:transits}) through
$|\sin\epsilon_\Gamma|$, is then
$\sin 2^\circ = 0.035$, a factor of five below the $0.17$ of the zero-order plate. The price is the
glass. A catalog unit carries some $43$\,mm of ZnSe per pass, so the round trip adds
$\approx 0.41$\,ns of group delay, nearly eight times the entire Nyquist window of
Table~\ref{tab:numbers} ($\approx 53$\,ps), and restoring $\dT$ means moving the idler
end mirror by ${\approx}6$\,cm. The rhomb is thus a dedicated chiral configuration,
whereas the zero-order plate, a fraction of a millimeter of material, is the
switchable version.

\paragraph{Polarimetric precision in context.}
The bounds of Sec.~\ref{sec:fim} are stated in the parameters of the Jones matrix,
whereas polarimetry speaks of retardance, diattenuation, and axis angles.
Table~\ref{tab:polarimetry} converts them for the example channel.
Its entries assume a low-gain source delivering of order $10^8$ detected photons per
second summed over both ports, so that $N = 10^6$ is a $10$\,ms exposure. A single
frame then resolves the round-trip retardance to $0.33^\circ$ and the diattenuation
to $0.55$ percentage points, and a one-second campaign ($N = 10^8$) brings these to
$0.033^\circ$ and $0.055$ percentage points (single-pass values are about half of the
round-trip ones). With Poisson noise alone all entries scale
as $1/\sqrt{N}$, so the table reads as a conversion between integration time and
precision for any rate; its last row is the mid-infrared dose on the sample that buys
this precision. Camera read noise breaks that scaling at the single-frame budget: with
$15\,e^-$ rms per pixel, the Gauss--Poisson variance $G\,I_P + \sigma_r^2$ of
App.~\ref{app:fim} at $G = 1$, i.e.,\ counts in photoelectrons, the $10$\,ms entries
rise by a factor $1.4$--$1.5$ (second column),
up to $1.5$ on the moduli. The reason is that the mean pixel then carries $170$
photons, averaged over the $771$--$831$\,nm grid of App.~\ref{app:fim}, against a
read-noise variance of $\sigma_r^2 = 225$ electrons$^2$;
the one-second campaign is unaffected to within $1\%$. The instrument function
of $20$\,pm and the crystal chirp are included in both columns.

\begin{table}[htb]
\caption{\label{tab:polarimetry}%
CRBs ($1\sigma$) of the single-frame readout in polarimetry
units, for the example channel of Fig.~\ref{fig:demo} (round-trip retarder
$0.8$\,rad at $30^\circ$, diattenuator $0.9/0.7$, rotation $20^\circ$) with the
$20$\,pm instrument function, the crystal chirp and $\Phz$ estimated alongside; first
column Poisson noise only, second column with $15\,e^-$ rms camera read noise per
pixel. All quantities refer to the measured round-trip
Jones matrix $M$; retardance, diattenuation, and the two axes follow the polar
decomposition of $M$ into a diattenuator followed by a retarder, and the optical
rotation is half the circular component of that retarder. The pair
$\Psi_{\rm ell}$, $\Delta_{\rm ell}$ is the transmission
analogue of the ellipsometric ratio in the H/V basis,
$\tan\Psi_{\rm ell}\,e^{i\Delta_{\rm ell}} = \Mhh/\Mvv$; $\Delta_{\rm ell}$ is not one
of the carrier delays $\Delta_f$, and pp = percentage points. The frame time
assumes $10^8$ detected photons per second summed over both ports, the order of
magnitude of our present source. The Poisson bounds scale as $1/\sqrt{N}$, so a one-second
campaign ($N = 10^8$) divides every entry of the first column by ten, and the read-noise
column then coincides with it to within $1\%$. The last row counts the mid-IR
photons sent into the sample arm: one half of the detected number at unit detection
efficiency, since only first-pass pairs probe the channel (Sec.~\ref{sec:rates}),
rising by the inverse detection efficiency below that.}
\begin{ruledtabular}
\begin{tabular}{lcc}
 & \multicolumn{2}{c}{$10$\,ms frame, $N = 10^6$} \\
 & Poisson & $+\,15\,e^-$ read noise \\
\hline
Total retardance & $0.33^\circ$ & $0.47^\circ$ \\
Retarder axis & $0.22^\circ$ & $0.31^\circ$ \\
Diattenuation $D$ & $0.55$\,pp & $0.80$\,pp \\
Diattenuator axis & $0.66^\circ$ & $0.94^\circ$ \\
Optical rotation & $0.17^\circ$ & $0.24^\circ$ \\
Amplitude ratio $\Psi_{\rm ell}$ & $0.18^\circ$ & $0.26^\circ$ \\
Phase difference $\Delta_{\rm ell}$ & $0.38^\circ$ & $0.54^\circ$ \\
$|M_{ij}|$, relative & $3$--$10\times10^{-3}$ & $5$--$14\times10^{-3}$ \\
\hline
Mid-IR photons on sample & \multicolumn{2}{c}{$5\times10^5$ ($26$\,fJ)} \\
\end{tabular}
\end{ruledtabular}
\end{table}
These numbers invite a comparison with classical instruments. Two of them come
close to the full complex Jones matrix: mid-IR dual-comb polarimetry in the fingerprint region
(near $8\ \mu$m) records amplitude and absolute phase for all four polarization
channels, one polarizer setting at a time, over up to $90$\,cm$^{-1}$ with
$65\ \mu$s per spectrum \cite{hinrichs2023}, and delay-multiplexed dual-comb polarimetry
recovers all eight real parameters of $M$, near $1560$\,nm \cite{koresawa2026}. Both
report the Jones elements without uncertainties or an information analysis, as does
single-input polarization-sensitive OCT, which spreads the input polarization over
the source spectrum and reads depth-resolved retardance and axis orientation from one
frame \cite{jones2026}. Polarimetry covering the
$3$--$5\ \mu$m band has been reported with intensity-based Mueller-matrix
ellipsometers \cite{garciacaurel2015,furchner2018,furchner2020}, which retrieve the
ellipsometric angles $\Psi_{\rm ell}$ and $\Delta_{\rm ell}$ of Table~\ref{tab:polarimetry}
from sequences of polarizer and retarder settings through a nonlinear inversion; a
field-level single-frame measurement of amplitude and phase per Jones channel has not
been reported there. The claim of Table~\ref{tab:priorart} is
therefore not spectral Jones-matrix polarimetry in the mid-IR as such, but its
single-frame version without a mid-IR detector and with the estimation theory
attached.

\paragraph{Relation to quantum polarimetry and SU(1,1) estimation theory.}
The estimation theory this work uses has two neighbors, neither of which covers
it. Fisher-information treatments of polarimetry with quantum light are well
developed, but all of them detect the light that probed the sample, and none of them
is set in the mid-IR: quantum limits
of ellipsometry \cite{rudnicki2020} and multiparameter bounds for chiral media
\cite{wang2021cd} are wavelength-agnostic theory; entangled two-channel polarimetry
has been analyzed and demonstrated at $810$\,nm \cite{pedram2024}, and two-parameter
polarimetry approaching the quantum Cram\'er--Rao bound has been demonstrated with
photon pairs at $1550$\,nm \cite{niblo2026}. See
Ref.~\cite{goldberg2021} for a review. Conversely, the estimation theory of SU(1,1)
interferometers counts both output modes and concerns phase and loss rather than
polarization \cite{yurke1986,marino2012}, and the estimation theory of sensing with
undetected photons does leave the probing beam undetected, but likewise treats only
one or two real parameters per sample \cite{panda2026,houde2026,zhang2025h}. The present analysis sits in the intersection the two literatures
leave open: polarization-channel estimation in a nonlinear interferometer
without detecting the probing beam. One result of Ref.~\cite{panda2026} bears directly
on the fold. In its scalar model the information on a phase after $n$ passes through
a sample of single-pass amplitude transmission $t_{\rm sp}$ grows as
$n^2 t_{\rm sp}^{2n}$, so the
optimal number of passes is $n^\ast = -1/\ln t_{\rm sp}$ [Eq.~(58) of Ref.~\cite{panda2026}],
and a fixed double pass, which is what the fold provides (its round-trip amplitude
being $t_{\rm sp}^2$), beats a single pass at equal
photon number whenever $t_{\rm sp} > 1/2$ and is optimal near $t_{\rm sp} = 0.6$. For the example
channel, whose single-pass amplitude transmissions are $0.95$ and $0.84$, the round
trip gains a factor $4t_{\rm sp}^2 = 2.8$--$3.6$ in the information on the \emph{single-pass}
phase over one pass. Samples
more transparent than that would profit from further passes, which the fold does not
offer.

\paragraph{Outlook.}
The immediate next step is the experiment itself: the design point of
Sec.~\ref{sec:feasibility} describes the source with which we intend to implement
the scheme. Beyond it, extension to
$d$-dimensional mode spaces (combining delay tags with the mode unitaries of
Ref.~\cite{rajeev2026}), adaptive working-point control (worth little for the
Jones parameters, Sec.~\ref{sec:working}, but a lever once a weighted,
application-specific criterion replaces $\det F$), and the high-gain route sketched above are open
directions. The estimation-theoretic treatment started here extends to each.

\section*{Data and code availability}
The scripts that generate all figures and tables of this paper will be deposited on
Zenodo; the DOI will be inserted at acceptance.
\begin{acknowledgments}
M.Z.\ gratefully acknowledges the hospitality of the group of Guo-Yong Xiang and
Zhibo Hou at the University of Science and Technology of China, Hefei, where the
idea for this work took shape during a one-week visit.
M.Z.\ thanks her AI assistant Klaus for countless working
sessions; the disclosure below states what it did and did not contribute.

\textbf{AI assistance.} The authors used Claude (Anthropic; Claude Opus 5 and
Claude Fable 5.1, accessed 2026) as an AI coding and writing
assistant during the preparation of this work. It was used to write and
cross-check symbolic derivations (SymPy) and numerical simulation scripts, to
assist with literature searches, to generate the rendering script for
Fig.~\ref{fig:setup}, and to edit the language of the manuscript. All analytical
results were independently re-derived, all numerical results were reproduced
independently, and all cited references were verified against the original
publications by the authors. The AI assistant is not an author and made no
independent scientific contribution; the authors take full responsibility for the
content of this paper.
\end{acknowledgments}

\appendix

\section{Detection rates from the full output state}
\label{app:rates}
Equations~(\ref{eq:master})--(\ref{eq:phases}) and Table~\ref{tab:families} follow
from a first-order (low-gain) amplitude calculation of the double-passed source
with the idler channel embedded in a vacuum-ancilla dilation [cf.\
Ref.~\cite{rajeev2026}, Eqs.~(5)--(7)]. We write out the Schr\"odinger-picture
chain. An independent Heisenberg (input--output) derivation and a numerical
dilation model agree on every gauge-invariant quantity.

\emph{Conventions.} The CW pump ties the detunings together: signal at
$\omega_{s0}{+}\Omega$ pairs with idler at $\omega_{i0}{-}\Omega$, with $\Omega$
the signal detuning from band center. One crystal
transit with polarization $\mu \in \{y, z\}$ contributes the phase
$\varphi^s_\mu = \alpha^s_\mu + \Omega\,\tau_\mu^s$ to the signal and
$\varphi^i_\mu = \alpha^i_\mu - \Omega\,\tau_\mu^i$ to the idler, with $\alpha$
the phase at band center and $\tau$ the group delay per transit. The derivation keeps the
linear order in $\Omega$, which fixes the carrier ladder; the group-velocity
dispersion within a transit is a nonlinear remainder $\varphi^{\rm nl}_\mu(\Omega)$
that rides on the same ladder and is treated in App.~\ref{app:transits},
Eq.~(\ref{eq:chirpladder}). The monochromatic pump acquires
the constants $\varphi^p_y$ and $\varphi^p_z$ per $y$ and $z$ transit (its $z$
component is the one that drives type-0 generation), and $\kappa$ is the relative
generation phase of Eq.~(\ref{eq:pass1}). Throughout we assume identical crystals,
equal gain for all four generation processes, and ideal dichroic mirrors, end
mirrors, and PBS. We also neglect the polarization dependence of the crystals' own
mid-IR absorption: toward the infrared edge of KTP the $y$ and $z$ transits of the
idler are attenuated differently, and since the $V$-probe idler crosses C2 in $y$
while the $H$-probe idler does not, a real pair of crystals adds a wavelength-dependent
instrument diattenuation in front of the channel, which we do not model here. Its scale
is set by the extra crystal transit rather than by the anisotropy itself: with an
absorption coefficient of about $0.4\,{\rm cm}^{-1}$ at band center, rising to about
$1.6\,{\rm cm}^{-1}$ at $4.2\ \mu$m, plus a narrow band at $3.47\ \mu$m
\cite{hansson2000}, one $2.55$\,mm transit costs the $V$ probe between roughly $5\%$ and
$20\%$ in amplitude across the readout window, a static instrument diattenuation that
the reference frame of Sec.~\ref{sec:recipe} absorbs into the instrument contrasts as long
as it is stable between frames. The
single-pass delay $T$ of the signal-arm plate is taken to have
equal phase and group delay (otherwise $\omega_{s0}T$ is replaced by the plate's
phase delay $\varphi_T$ in the calibration phases $\eta_f$).

\emph{(i) Pass 1.} With $\chi$ the pump polarization angle, the pump component
$\cos\chi$ transits C1 as $y$ and generates
the $H$ pair in C2 (facing the arms). The component $e^{i\kappa}\sin\chi$
generates the $V$ pair in C1, which then transits C2 with both photons
$y$-polarized. The two probe amplitudes therefore carry the prefactors $p_H$ and $p_V$,
\begin{equation}
p_H = \cos\chi\; e^{i\varphi^p_y}, \qquad
p_V = e^{i\kappa}\sin\chi\; e^{i(\varphi^s_y + \varphi^i_y)} .
\label{eq:prefactors}
\end{equation}

\emph{(ii) Arms.} The signal component of probe $j \in \{H, V\}$ returns as
column $j$ of the double-passed plate--wave-plate operator
$J_{\rm sig} = \mathcal{P}(T)\,R(A)\,\mathcal{P}(T)$ (App.~\ref{app:transits}), where $R(A)$ is the
Jones matrix of the signal wave plate, written as a reflection
(Sec.~\ref{sec:folded}). The idler component
returns through the dilated channel, feeding $M_{kj}$ into idler mode $k$ and
$C_{lj}$ into ancilla mode $l$ with $C^\dagger C = \openone - M^\dagger M$. Both
photons share the round-trip factor
$u = \exp i\big[(\omega_{s0}{+}\Omega)\tau_s + (\omega_{i0}{-}\Omega)\tau_i +
\Xi\big]$, where $\tau_s$ and $\tau_i$ are the round-trip delays of the signal and
idler arms, not the per-transit delays $\tau_\mu^{s,i}$. The term
$\Xi = \varphi^s_y + \varphi^s_z + \varphi^i_y + \varphi^i_z$ is the return
double transit through both crossed crystals. It is polarization independent for
equal lengths, since $\mathcal T_{\rm C2}\mathcal T_{\rm C1} \propto \openone$ with
$\mathcal T_{\rm C1}$, $\mathcal T_{\rm C2}$ the transit operators of the two
crystals. That is why the return creates no new families.

\emph{(iii) Pass 2.} By path identity the returning pump, $(b_H,\, b_V
e^{i\theta})$ at the arm-side face of C2, adds the reference amplitudes
coherently. For the bare pump mirror of
Fig.~\ref{fig:setup} it is simply $(\cos\chi,\, \sin\chi)$. Here $\theta$ collects
the phase asymmetry of the return path (mirror, dichroics) between the two
pump components, together with the input pump's own $H$--$V$ phase, which a bare
mirror hands on unchanged. It stays a calibration constant and enters the
observables only through the combination $\theta + \kappa$. On its forward
pass each pump component has transited both crystals, once as $y$ and once as
$z$. With $\omega_p$ the pump frequency, the pump-arm phase
$v_p = e^{i\omega_p \tau_p}$ counts the arm alone, from
the arm-side face of C2 to the mirror and back. The references are therefore
$b_H\, v_p\, e^{i(\varphi^p_y + \varphi^p_z)}\, e^{i(\varphi^s_y + \varphi^i_y)}$
on the $(H_s, H_i)$ mode and
$b_V\, v_p\, e^{i(\theta + 2\varphi^p_y + \varphi^p_z)}$ on $(V_s, V_i)$.
For the $(H_s, H_i)$ term the pump transits C1 as $y$ and C2 as $z$, generation is
in C2, and the pair transits C1 backward as $y$. For the $(V_s, V_i)$ term the pump
transits C1 as $z$ and C2 as $y$ and returns through C2 as $y$, with generation in C1.

\emph{(iv) Port projection.} The rate at PBS port $P$ is the modulus squared,
summed over all undetected modes (idler polarizations and
ancillas),
\begin{equation}
I_P \propto \sum_k \big|\psi_{Pk}\big|^2
+ \sum_{jj'} a^{*}_{Pj}\, a_{Pj'} \big(\openone - M^\dagger M\big)_{jj'} ,
\label{eq:portrate}
\end{equation}
with $\psi_{Pk}$ the detected-idler amplitudes assembled from (i)--(iii) and
$a_{Pj} = p_j\, u\, (J_{\rm sig})_{Pj}$ the probe amplitudes entering the ancilla
sum, the $p_j$ being those of Eq.~(\ref{eq:prefactors}). Expanding
Eq.~(\ref{eq:portrate}) term by term reproduces
Eqs.~(\ref{eq:master})--(\ref{eq:phases}) with the weights, carriers, and flags for
the global interferometer phase $\Phz$ listed in Table~\ref{tab:families}. The
ladder origin $\Delta_1$ and the rung $d_1$, in terms of the signal--idler
group-delay imbalance $\dT$ and the single-pass plate delay $T$, are
\begin{equation}
\Delta_1 = \dT + (\tau_z^s - \tau_z^i), \qquad d_1 = T + (\tau_y^s - \tau_y^i) .
\label{eq:ladder}
\end{equation}
Two calibration phases appear explicitly,
\begin{equation}
\eta_1 = \alpha^s_z + \alpha^i_z - \varphi^p_z ,
\label{eq:eta1}
\end{equation}
\begin{equation}
\varepsilon = \kappa + \omega_{s0}T + \alpha^s_y + \alpha^i_y - \varphi^p_y ;
\label{eq:epsilon}
\end{equation}
they enter the ladder relations of result (b) below. In this bookkeeping
$\eta_1 = \big[k_z(\omega_{s0}) + k_z(\omega_{i0}) - k_z(\omega_p)\big]\,L$ is the
$z$-axis phase mismatch of one crystal transit, with $L$ the crystal length. It
vanishes at perfect phase matching and reduces to the grating phase,
$\eta_1 = -2\pi L/\Lambda_{\rm QPM}$ (mod $2\pi$), with $\Lambda_{\rm QPM}$ the
poling period, for a quasi-phase-matched
crystal. (Absorbing the pump's forward transits into $\tau_p$ instead moves
$\varphi^p_y + \varphi^p_z$ from $\Phz$ into $\eta_1$. The phase $\varepsilon$ and the
ladder relations are the same in either convention.)

Three structural results follow.
(a)~\emph{Offsets are exactly channel independent}: each probe column
contributes $(M^\dagger M + C^\dagger C)_{jj} = 1$, so idler loss never changes the
singles rate, only the fringe contrast. The bookkeeping closes only once the ancilla
modes are summed.
(b)~\emph{Carrier and phase ladder}: carriers form the equidistant ladder
$\Delta_1,\ \Delta_1 + d_1,\ \Delta_1 + 2d_1$ counting the delay-plate passes
[Eq.~(\ref{eq:ladder})], so that $\Delta_2 = \Delta_3 = \Delta_1 + d_1$ and
$\Delta_4 = \Delta_1 + 2d_1$, and the constant phases follow the same ladder,
$\eta_2 = \eta_1 + \varepsilon$, $\eta_4 = \eta_3 + \varepsilon$,
$\eta_3 = \eta_2 - (\theta + \kappa)$,
with $\eta_1$ and $\varepsilon$ from Eqs.~(\ref{eq:eta1}) and (\ref{eq:epsilon}). Only three independent
calibration constants exist; a pump $H$--$V$ phase, e.g.,\ a circularly instead of
diagonally polarized pump, only shifts them, since the pump ellipticity phase is
gauge and only the amplitude ratio $\chi$ enters the weights.
(c)~\emph{The dark carrier}: the probe--probe amplitude is weighted by
$(M^\dagger M + C^\dagger C)_{HV} = \openone_{HV} = 0$. Keeping only $M$ would
predict a spurious $\Phz$-free fringe $\propto \dephw$ at carrier $d_1$, of
opposite sign in the two ports. That is a useful diagnostic for an incorrect loss
treatment, and the reason Sec.~\ref{sec:witness} can use this carrier as a null
test.

\section{Transit bookkeeping and delay budget}
\label{app:transits}

This appendix collects the delay bookkeeping behind the carrier ladder: the
splitting the crystals supply by themselves, the delay plate that sets $d_1$
(Sec.~\ref{sec:walkoff}), and the chirp each family inherits
(Sec.~\ref{sec:chirp}). It then draws two consequences of the degenerate middle
rung, the PBS ghost of Sec.~\ref{sec:placement} and the swap-plate calibration of
Sec.~\ref{sec:discussion}.

\emph{Natural splittings.} For $\chi \neq 0$ the $V$ pair traverses C2 with both
photons $y$-polarized (see App.~\ref{app:rates}, step (i)), so its natural carrier
offset relative to the $H$ pair
is $(L/c)[n_g^y(\omega_{s0}) - n_g^y(\omega_{i0})] = \delta L$. This is the same
walk-off as the reference split, Eq.~(\ref{eq:delta}). With $\delta = -0.302$\,ps/mm
(App.~\ref{app:sellmeier}) the natural $d_1$ is $-0.77$\,ps for our $2.55$\,mm
crystals. That is far too small once each carrier has to support a $\pm2.5$\,ps
demodulation window, so an engineered delay is required.

\emph{Double-passed delay plate.} A birefringent plate in the signal arm, before
the wave plate (where the two first-pass signal components are still orthogonally
polarized), is traversed twice. With a single-pass delay $T$, the composite Jones
operator is
\begin{equation}
J_{\rm sig} = \mathcal{P}(T)\, R(A)\, \mathcal{P}(T), \qquad \mathcal{P}(T) = \mathrm{diag}(1, e^{i\Omega T}),
\end{equation}
which evaluates to
\begin{equation}
J_{\rm sig} = \begin{pmatrix} \cos A & \sin A\, e^{i\Omega T} \\
\sin A\, e^{i\Omega T} & -\cos A\, e^{2i\Omega T} \end{pmatrix} ,
\end{equation}
i.e.,\ element delays $(0, T, T, 2T)$: the per-port carrier spacing is $T$, the
$VV$ element appears at $2T$, and the dark carrier sits at $T$. Its input half is
the passive polarization-delay unit of classical Jones-matrix OCT
\cite{baumann2012,lim2012,ju2013}, where the delay is applied on illumination only
and the two detection channels are separated by a polarizing beam splitter,
polarization-diversity detection \cite{lim2012}, giving
element delays $(0,T,0,T)$. Double-passing the plate here adds the matching delay on
the detection side; the doubled element $VV$ and the dark carrier at $T$ have no
counterpart there. The
carrier placement of Table~\ref{tab:numbers} corresponds to $d_1 = 5$\,ps
(F1 at $\dT = 15$\,ps, F2/F3 at $20$\,ps, F4 at $25$\,ps, dark carrier at $5$\,ps,
Nyquist at $53$\,ps). The analytic derivation (App.~\ref{app:rates}) shows that the
crystal transits do not disturb this structure. They renormalize the rung spacing,
so the plate is specified at $T = 5.77$\,ps in order to leave the nominal carrier
splitting $d_1 = T + (\tau_y^s - \tau_y^i) = 5.77 - 0.77 = 5.0$\,ps net of the
crystal walk-off ($-0.302$\,ps/mm over our $2.55$\,mm crystals). They also
shift the whole ladder by $\tau_z^s - \tau_z^i = +33$\,fs. The quoted
placements are therefore nominal. In practice the demodulation windows are
centered on the renormalized carriers, which are read directly off the frame. In the $\chi = 0$, $T = 0$ limit the first rung reduces to
$\tau_y^s - \tau_y^i = \delta L$, the reference walk-off of the state limit.

\emph{Plate design.} $T = 5.77$\,ps requires a group-delay path difference
$c\,T \approx 1.73$\,mm, i.e.,\ a plate length
$\ell_{\rm plate} = c\,T/|\Delta n_g|$, with $\Delta n_g$ the group birefringence
of the plate at $800$\,nm. That is $9.5$\,mm of calcite, $7.2$\,mm of YVO$_4$, or
$16$\,mm of a KTP-like material; quartz is impractical at $180$\,mm.
Calcite is the natural choice: the shortest practical plate with the least
differential dispersion. Across the readout window the plate chirp alone broadens
the carriers by at most a few delay bins ($0.12$\,ps on the $2T$ rung for the
worst material, a KTP-like plate; $0.07$\,ps for calcite), a factor $40$ to $70$
below the $5$\,ps carrier spacing. The plate sits in the signal arm and only ever sees $800$\,nm, so there
is no mid-IR transparency constraint, one more dividend of the three-arm geometry.
The plate chirp is, however, the small part of the story; the crystal transits of
the idler contribute the large part, treated next.

\emph{Crystal chirp ladder.} The conventions of App.~\ref{app:rates} keep each transit
phase to linear order in $\Omega$. Let $\varphi^{\rm nl}_\mu(\Omega)$ denote the
remainder beyond that order for one signal--idler transit pair in polarization $\mu$,
\begin{equation}
\begin{split}
\varphi^{\rm nl}_\mu(\Omega) = {}& L\Big[k_\mu(\omega_{s0}{+}\Omega)
+ k_\mu(\omega_{i0}{-}\Omega)\Big] \\
& - \alpha^s_\mu - \alpha^i_\mu - \Omega\,(\tau_\mu^s - \tau_\mu^i) ,
\end{split}
\label{eq:chirpladder}
\end{equation}
i.e.,\ the transit-pair phase after removing the band-center phase and the net group
delay of App.~\ref{app:rates}, both of which are already booked in the ladder. Let
$\varphi^{\rm nl}_T$ denote the same remainder for the delay
plate (signal only, per single pass; the double pass is what puts $2\varphi^{\rm nl}_T$
on F4). The same bookkeeping, steps (i)--(iv) of App.~\ref{app:rates}, assigns these
remainders to the families on the same ladder as the delays (Sec.~\ref{sec:chirp}).
With the Sellmeier model of App.~\ref{app:sellmeier}
over $782$--$816$\,nm the ladder increment
$\varphi^{\rm nl}_y + \varphi^{\rm nl}_T$ carries a group-delay dispersion of
$-1.2\times10^{4}$\,fs$^2$ and a third-order term of $-2.0\times10^{5}$\,fs$^3$.
The idler transit in $y$ dominates, with $-1.3\times10^{4}$\,fs$^2$ near
$3.8\ \mu$m, $94\%$ of the three contributions by magnitude; the signal transit at
$800$\,nm adds $+5\times10^{2}$\,fs$^2$ and the calcite plate
$+3\times10^{2}$\,fs$^2$. The common term $\varphi^{\rm nl}_z$
carries $-3.2\times10^{3}$\,fs$^2$ (idler $-3.9\times10^{3}$, signal $+7\times10^{2}$).
Group delays across the window thus spread by $0.36$, $1.9$, and $3.4$\,ps on F1,
F2/F3, and F4. The consequences for demodulation, resolution, and guard bands are
drawn in Sec.~\ref{sec:chirp}.

\emph{Ghost peaks.} With the symmetric $(0, T, T, 2T)$ structure, F2 ($D_H$) and
F3 ($D_V$) sit at exactly the same carrier. The ladder is degenerate on
its middle rung, and the two families are separated by polarization alone, so finite
PBS extinction maps each onto the other port \emph{on top of} the genuine peak. The
leak keeps its polarization, orthogonal to the light that belongs on that port, so
the two add in intensity. The ghost is the neighboring family's fringe scaled by the
leak fraction $\epsilon_{\rm PBS}$ itself, not by $\sqrt{\epsilon_{\rm PBS}}$, and the
bias on the coincident modulus is $\epsilon_{\rm PBS}$ times the ratio of the two
fringe amplitudes. For the transmitted port of a standard cube
($\epsilon_{\rm PBS} \approx 10^{-3}$) this is a $0.1\%$ effect on $|\Mhv|$. The
reflected port of a broadband cube, however, is typically specified at only
$100{:}1$ down to $20{:}1$, i.e.,\ $1$ to $5\%$ on $|\Mvh|$ alone. That is an asymmetric error
that lands directly on the ratio $|\Mhv|/|\Mvh|$ and hence on the reciprocity check
of Sec.~\ref{sec:demo}; it is two orders of magnitude above the $10^{-3}$ accuracy
the moduli otherwise reach.

Two symmetry-breaking
candidates fail or are impractical. A birefringent
plate in the \emph{pump} arm does not help. The pump is monochromatic, so
a $V$-pump delay $T'$ is a pure phase that shifts
$\theta \to \theta + 2\omega_p T'$ and no carrier (verified within the
formalism of App.~\ref{app:rates}).
Deliberately unequal crystal lengths $L_1$ (C1) and $L_2$ (C2) do lift the
degeneracy. The derivation generalizes, the shift is rigid per port, and no new
families appear; the splitting is
$\Delta_3 - \Delta_2 = (L_1 - L_2)\,[n_g^z(\omega_{s0}) - n_g^z(\omega_{i0})]/c$. But the scale
is the $z$-axis walk-off of $13$\,fs/mm (App.~\ref{app:sellmeier}): a useful
separation of a few hundred fs would take tens of mm of length difference. The
practical answer, and the one we adopt, is a Glan--Thompson clean-up polarizer
(extinction $\gtrsim 50$\,dB) behind \emph{both} PBS ports, so that
the two traces are attenuated symmetrically. The residual leak then enters
coherently, with amplitude
$\sqrt{\epsilon_{\rm PBS}\,\epsilon_{\rm clean}}$, where $\epsilon_{\rm clean}$ is
the leak fraction of the clean-up polarizer. That is $3\times10^{-4}$ for a
$100{:}1$ PBS and a $50$\,dB polarizer, below the $10^{-3}$ accuracy the moduli
otherwise reach. The angle of the clean-up polarizer then becomes the dominant term, one that is
open to calibration. The leak is moreover \emph{deterministic}, since F3's complex amplitude
is measured on its own port. A calibrated leak (amplitude and phase, once the
clean-up polarizer has made it coherent) can therefore be subtracted in
postprocessing.

\emph{The swap plate as a calibration switch.} The calibrated $\epsilon_{\rm PBS}$
this recipe requires can be measured in situ, with the same swap plate that
Sec.~\ref{sec:discussion} uses for the chiral mode. The flip-in quarter-wave plate at
$45^\circ$ in front of the idler end mirror acts in double pass as the exact
exchange $\sigma_x$. This is the same operator as the signal-arm reflection $R(A)$
at $A = 90^\circ$, with no residual phase, so an empty idler arm returns
$M = \sigma_x$. Then F1/F4 ($\Mhh$, $\Mvv$) go dark and F2/F3 ($\Mhv$, $\Mvh$)
reach full amplitude, the exact complement of the swap-free configuration
(Table~\ref{tab:families}). Setting additionally $B = 0$ (returning pump purely
$H$, $b_V = 0$) switches off F3 and F4, leaving F2 as the \emph{only} live
family on the degenerate carrier. Whatever appears at $\dT + d_1$ on the $D_V$
port \emph{is} the PBS ghost, read out in amplitude and phase in one frame.
This $B = 0$ is realized by a flip-in quarter-wave plate at $22.5^\circ$ in
front of the pump end mirror, whose double pass reflects the diagonal pump onto
pure $H$ at unit amplitude. That is the very map that forbids a permanent plate there
(Sec.~\ref{sec:working}), put to use as a calibration switch.
Without the swap this calibration is inaccessible --- F2/F3 are dark for an
empty arm --- and it targets precisely the one readout violation the
null test cannot see (Sec.~\ref{sec:witness}). The dark carrier at $d_1$ stays
dark in either mode, since its weight $(M^\dagger M + C^\dagger C)_{HV} =
\openone_{HV} = 0$ holds for arbitrary $M$ and $\sigma_x$ is unitary. The
crossed-crystal null test of Sec.~\ref{sec:witness} runs
unchanged, and the two null tests are independent.

Finally, the empty-arm
F1/F4 residuals diagnose the swap plate itself at first order. An angle error
$\epsilon_{\rm sw}$ gives $\Mhh = -2\epsilon_{\rm sw}$, real and antisymmetric,
$\Mhh = -\Mvv$. A plate error of $1^\circ$ produces a fringe amplitude of $3.5\%$,
comfortably measurable. A retardance error $\epsilon_{\Gamma}$ gives $\Mhh = -i\epsilon_{\Gamma}$
(imaginary and \emph{symmetric}, $\Mhh = +\Mvv$). It is the only symmetric
perturbation in the error budget, hence isolated by the sum F1${}+{}$F4, and
separated from the angle error by phase. Sample responses land on the same
diagonal with their own signatures (a reciprocal rotation $\alpha_{\rm OA}$ real and
antisymmetric, circular dichroism $\alpha_{\rm CD}$ imaginary and antisymmetric, in
quadrature). The consequences for the chiral operating mode are discussed in
Sec.~\ref{sec:discussion}.

\section{Sellmeier numerics for ppKTP}
\label{app:sellmeier}

All dispersion numbers in the main text derive from one consistent model:
$n_z$ from Katz \emph{et al.}~\cite{katz2001,katz2002}, a room-temperature Sellmeier
equation determined there for the entire transparency range of flux-grown KTP and
used here for the idler; $n_y$ and $n_z$ in the visible
and near infrared from Fan \emph{et al.}~\cite{fan1987}, used for the pump and the
signal; $n_y$ at the idler from Kato--Takaoka~\cite{kato2002}, the only published $n_y$ fit
we are aware of that reaches $3.54\ \mu$m; and temperature derivatives from
Emanueli--Arie~\cite{emanueli2003}. The poling period is expanded thermally along
$x$ with the coefficients of Smith \emph{et al.}~\cite{smith2016}
($\alpha^{\rm th}_x = 7.88\times10^{-6}$/K at $305$\,K).
At $25\,^\circ$C and the nominal $800$\,nm working point, the model gives
retracing zero crossings at $787.3$ and $807.6$\,nm and a group-matched center
at $796.7$\,nm $\leftrightarrow 3.846\ \mu$m. The center tuning is $-31$\,pm/K for
the signal and $+0.72$\,nm/K for the idler. The tuning rates are the slope of a nearly flat curve at
its stationary point and are correspondingly model sensitive. The model gives a reference walk-off [Eq.~(\ref{eq:delta})]
$\delta = [n_g^y(\omega_{s0}) - n_g^y(\omega_{i0})]/c = -302$\,fs/mm ($-0.302$\,ps/mm in the
rounding of the main text) and a birefringent phase drift
[Eq.~(\ref{eq:phibir})] of $-16$\,mrad/(K\,mm). The signal--idler
walk-off in $z$ is $+13$\,fs/mm, and the $y$--$z$ signal walk-off is $-352$\,fs/mm
(the lever for the $d_1$ delay plate).
The crystal length is a trade against QPM retracing: the zero-mismatch curve is
length independent, while the length sets only the width of the sinc envelope around
it [cf.\ Eqs.~(1)--(2) of Ref.~\cite{vanselow2019}]. Our $L = 2.55$\,mm sits at the useful end of that trade. In the
model, the mismatch curve stays close to zero, so the efficiency
$\mathrm{sinc}^2(\Delta k L/2)$ stays a single hump: it dips to $0.80$ between the two
crossings, peaks at $808$\,nm, and has its FWHM at $782$--$816$\,nm ($34$\,nm) at our
length. At $L = 5$\,mm the same dip falls to $0.40$, so the length trade is real. The
depth of that dip is not a robust prediction: a refractive-index error of
$10^{-4}$ in the Sellmeier fit shifts $\Delta k$ by ${\sim}10^{-3}\ \mu$m$^{-1}$, comparable
to the whole excursion of the retracing curve. With the $n_z$ of Fradkin
\emph{et al.}~\cite{fradkin1999} in place of Katz \emph{et al.} for pump, signal
and idler alike, for instance, the
excursion reaches $-1.5\times10^{-3}\ \mu$m$^{-1}$ instead of
$-0.6\times10^{-3}\ \mu$m$^{-1}$, the dip falls to $0.23$, and the envelope splits into
two lobes. The source in the laboratory shows a single hump, consistent with the model of
Katz \emph{et al.}
Two caveats: the $n_y$ fit ends at $3.54\ \mu$m, and the $dn/dT$ fit at
$1.585\ \mu$m (order of magnitude only in the mid-IR). The first is mild for the
group delays but matters for the chirp: the $y$-transit dispersion of
App.~\ref{app:transits} at the idler wavelengths is an extrapolation of the
Kato--Takaoka fit beyond its range. Sign and order of magnitude are robust, but
the value is not. The $z$ transit, which needs no extrapolation, gives the same
picture, with $-3.9\times10^{3}$\,fs$^2$ on the idler against
$+7\times10^{2}$\,fs$^2$ on the signal. This is one more reason to take the chirp from the
calibration frame rather than from the model (Sec.~\ref{sec:chirp}). The choice of
$n_z$ itself carries the same kind of spread: Katz \emph{et al.} and Fradkin
\emph{et al.}~\cite{fradkin1999} differ by $2.5\times10^{-3}$ in the idler group
index, which moves the $z$ walk-off from $+13$\,fs/mm here to $+21$\,fs/mm with
Fradkin \emph{et al.} Every $z$-transit number above (crossings, group-matched
center, ladder shift) is therefore model dependent at that level.

\section{Fisher information details}
\label{app:fim}

This appendix spells out the noise model and the marginalization behind the
$\det F$ of Sec.~\ref{sec:fim}, the sequential baseline of
Sec.~\ref{sec:seqvsmux}, and the QFI construction of Sec.~\ref{sec:qfi}.

\emph{Noise model and FIM.} Each pixel $j$ of port $P$ reports a Gaussian count
with mean $\propto I_P(\Omega_j;\boldsymbol\vartheta)$ [Eq.~(\ref{eq:master})] and
variance $G\,I_P(\Omega_j) + \sigma_r^2$, with $G$ the camera gain and $\sigma_r$
the read noise. For
$\sigma_r = 0$ and $G = 1$ this reproduces the Poisson information. The numerics
use the phase-matching envelope of App.~\ref{app:sellmeier} as the SPDC spectrum
$S(\Omega)$. The grid spans $771$--$831$\,nm, wide enough to let the envelope run
out, with $N_{\rm pix} = 3000$ pixels per trace at $20$\,pm. The ratios quoted in
Sec.~\ref{sec:fim} do not depend on the envelope shape: a Gaussian of comparable
width on the same grid changes them by less than $0.1\%$ (the read-noise ratio by a
few percent). Dropping the
subleading variance-derivative term, the information matrix is
\begin{equation}
F_{ab} = \sum_{P,j}
\frac{\partial_a I_P(\Omega_j)\; \partial_b I_P(\Omega_j)}{G\,I_P(\Omega_j) + \sigma_r^2} ,
\label{eq:fim}
\end{equation}
over the seven channel parameters augmented by the nuisance parameters. These are
the global phase $\Phz$; the calibration phases $\eta_f$, where these are not
independently calibrated; and, for the sequential baseline, the inter-frame drift
$\epsilon_{\rm d}$. All numbers quoted in the main text treat the $\eta_f$ as known
from the calibration frame and marginalize $\Phz$ alone; a finite calibration budget
adds its own variance to the three relative phases.
The Gaussian prior on $\epsilon_{\rm d}$ enters as a diagonal
information term. The channel covariance is the $7\times7$
block of $F^{-1}$, and ``$\det F$'' in the main text is the effective determinant
$\det(\,[F^{-1}]_{\rm channel}\,)^{-1}$ after this marginalization. All quoted
results are $\Phz$-independent (checked numerically).
Because the design knobs $(\chi, A, B)$ enter Eq.~(\ref{eq:master}) only through
the weights $w_f$, working-point optimization is a design problem over the weight
simplex. The determinant criterion is the natural one for a parameter vector that
mixes dimensionless moduli with phases in radians: a reparametrization multiplies
$\det F$ by a factor that does not depend on the design, so the ranking of working
points is unchanged. The balanced-point result quoted in Sec.~\ref{sec:working} is a direct grid
evaluation of Eq.~(\ref{eq:fim}) over single working points $(\chi, A, B)$; we did
not search over mixtures of working points, so the Kiefer--Wolfowitz equivalence
theorem~\cite{kiefer1960} is not invoked as a certificate. The marginalization above
follows the standard treatment of nuisance parameters in polarimetric design
\cite{foreman2008}. Carrier placement, by contrast, enters through
separability (Sec.~\ref{sec:readout}), not through Eq.~(\ref{eq:fim}) at fixed
resolution.

\emph{Sequential baseline used in the comparison.} The two state-mode frames of
Sec.~\ref{sec:seqvsmux} ($\chi = 0$ and $\chi = 90^\circ$, half the exposure each)
are treated as independent. The drift prior enters as a diagonal information term
$1/\sigma_{\rm drift}^2$ on the second frame's phase offset $\epsilon_{\rm d}$,
which is marginalized together with $\Phz$.

\begin{table}[t]
\caption{\label{tab:perparam}%
Per-parameter ratios of Cram\'er--Rao uncertainties between the three levels of
Eq.~(\ref{eq:hierarchy}), example channel, $\Phz$ marginalized; $F_{\rm sig+idl}$
in the bin-wise form. The geometric means over the seven parameters, $1.48$ and
$2.16$, are the inverse square roots of the determinant ratios, $0.45^{-1/2} = 1.49$
and $0.19^{-1/2} = 2.29$, up to the correlations that the determinant
includes and the diagonal does not.}
\begin{ruledtabular}
\begin{tabular}{lcc}
 & $\sigma_{\rm spec}/\sigma_{\rm sig}$ & $\sigma_{\rm sig}/\sigma_{\rm sig+idl}$ \\
\hline
$|\Mhh|$ & $1.61$ & $2.72$ \\
$|\Mhv|$ & $1.44$ & $2.24$ \\
$|\Mvh|$ & $1.49$ & $2.02$ \\
$|\Mvv|$ & $1.54$ & $2.22$ \\
$\arg\Mhv$ & $1.39$ & $2.21$ \\
$\arg\Mvh$ & $1.48$ & $1.99$ \\
$\arg\Mvv$ & $1.41$ & $1.82$ \\
\end{tabular}
\end{ruledtabular}
\end{table}
\emph{The $\chi = 0$ state limit as a cross-check.} At $\chi = 0$ the master form
collapses to the two state fringes of Eq.~(\ref{eq:staterates}). The four
process-only directions ($|\Mhv|$, $|\Mvv|$ and their phases) carry information
$\propto \sin^2\!\chi$ and therefore drop out of the
Fisher matrix. On the surviving block the
seven-parameter machinery reproduces the direct three-parameter state FIM (and
the corresponding QFI) to numerical precision under the reparametrization
$|\Mhh| = t_H\sqrt{\Pi_H}$, $|\Mvh| = t_V \purity \sqrt{1-\Pi_H}$,
$\arg \Mvh = \xi$. Here $t_H$ and $t_V$ are the calibrated transmissions and
$(\Pi_H, \purity, \xi)$ the state parameters of Sec.~\ref{sec:diagonal}. Against the one-fringe-per-setting protocol of
Ref.~\cite{fuenzalida2022b} at equal budget, the folded state frame gains exactly
$3/2$ on both visibility parameters, because it recovers the third of the photons
that idles on their non-fringing detector. On the fringe-phase difference $\xi$ the comparison is state
dependent, with ratios of $0.9$, $1.1$, and $1.3$ for the three benchmark states
$(\Pi_H, \purity, \xi) = (0.5, 1, 0)$, $(0.5, 0.7, \pi/2)$, and $(0.8, 1, \pi/4)$.
There the scanned phase is calibrated, whereas our $\Phz$
is marginalized. Combined with
the visibility gain, this gives $\det F$ ratios of $2.0$, $2.5$, and $2.9$ over the
same states. A parallel readout \emph{within} the two-source geometry of
Ref.~\cite{fuenzalida2022b} would gain a
factor of $2.1$--$2.4$.

\emph{QFI computation (Sec.~\ref{sec:qfi}).} The channel is embedded in a
$4\times4$ unitary dilation, giving per frequency bin the amplitude matrix
$\mathcal{A}(\Omega)$ (signal port $\times$ \{idler $H$, $V$, two ancillas\}). Then
$\rho_{\rm sig} = \mathcal{A}\mathcal{A}^\dagger$, while for signal${+}$idler the
one-idler-photon and
idler-vacuum sectors are block diagonal after tracing the ancillas. The SLD QFI of
each block is evaluated in its eigenbasis; this construction yields the hierarchy of
Eq.~(\ref{eq:hierarchy}). For the signal alone the frequency bins are genuinely
incoherent, since tracing the idler removes the phase between them. The
idler-inclusive state, by contrast, retains the coherence between bins, and treating
its bins independently amounts to dephasing them in $\Omega$. Because that dephasing
commutes with the parameter encoding, the bin-wise value is a lower bound on
$F_{\rm sig+idl}$; the SLD QFI of the fully coherent purified state (idler, ancillas
and all bins kept) is an upper bound. For the example channel the two give
$(\det F)^{1/7} = 7.5\times10^{7}$ and $8.4\times10^{7}$, against $1.47\times10^{7}$
for the signal alone, hence the range $2.3$--$2.4$ quoted in Sec.~\ref{sec:qfi}.
Table~\ref{tab:perparam} resolves the two ratios of Eq.~(\ref{eq:hierarchy}) per
parameter, as ratios of the Cram\'er--Rao uncertainties $\sigma_a = \sqrt{[F^{-1}]_{aa}}$
of the seven channel parameters with $\Phz$ marginalized.
Total flux is parameter independent
(verified to $10^{-6}$), so the per-pair QFI scales linearly with $N$. The QFI
implementation and the rate model behind Fig.~\ref{fig:detf} agree to a relative
deviation below $10^{-13}$.

\bibliography{references}

\end{document}